\documentclass[twocolumn]{aastex631}

\usepackage{amsmath}
\newcommand{\mrate}{ M$_{\odot} $yr$^{-1}$}
\newcommand{\halpha}{$\rm H\alpha$ }
\newcommand{\zsun}{$\rm Z_\odot$}
\defcitealias{tremblay_multiphase_2012}{T12a}
\defcitealias{tremblay_residual_2012}{T12b}
\defcitealias{mcnamara_discovery_2001}{M01}
\defcitealias{clarke_low-frequency_2005}{C05}

\usepackage[utf8]{inputenc}
\usepackage[T1]{fontenc}
\usepackage{textcomp}

\usepackage{graphicx} 
\usepackage{tabularx} 

\newcommand{\msun}{$M_\odot$}
\newcommand{\kms}{km s$^{-1}$}
\newcommand{\yr}{yr$^{-1}$ }

\newcommand{\chandra}{\textit{Chandra}}

\newcommand{\ovi}{{O VI} $\lambda$1032}

\usepackage{float}
\usepackage{enumitem}
\usepackage{natbib}
\DeclareUnicodeCharacter{2212}{-}

\begin{document}

\title{A Deep \textit{Chandra} View of Abell 2597: Bubbles, Shocks, Cold Fueling, and a Plasma Depletion Layer}

\author[0000-0002-3649-5362]{Osase Omoruyi}
\affiliation{Center for Astrophysics $|$ Harvard \& Smithsonian, 60 Garden St.,
Cambridge, MA 02138, USA}

\author[0000-0002-5445-5401]{Grant Tremblay}
\affiliation{Center for Astrophysics $|$ Harvard \& Smithsonian, 60 Garden St.,
Cambridge, MA 02138, USA}

\author[0000-0002-4735-8224]{Stefi A. Baum}
\affiliation{University of Manitoba, Dept. of Physics and Astronomy, Winnipeg, MB R3T 2N2, Canada}

\author[0000-0001-6812-7938]{Tracy E. Clarke}
\affiliation{U.S. Naval Research Laboratory, 4555 Overlook Avenue SW, Washington, DC 20375, USA}

\author[0000-0001-9212-3574]{Pratik Dabhade}
\affiliation{Astrophysics Division, National Centre for Nuclear Research, Pasteura 7, 02-093 Warsaw, Poland}

\author[0000-0002-9378-4072]{Andrew Fabian}
\affiliation{Institute of Astronomy, University of Cambridge, Madingley Road, Cambridge CB3 0HA, UK}

\author[0000-0003-2754-9258]{Massimo Gaspari}
\affiliation{Department of Physics, Informatics \& Mathematics, University of Modena \& Reggio Emilia, 41125 Modena, Italy}

\author[0000-0003-3785-1725]{Sanna Gulati}
\affiliation{National Centre for Radio Astrophysics, Tata Institute for Fundamental Research,
Savitribai Phule Pune University, Ganeshkhind, Pune 411007, India}

\author[0000-0003-3203-1613]{Preeti Kharb}
\affiliation{National Centre for Radio Astrophysics, Tata Institute for Fundamental Research,
Savitribai Phule Pune University, Ganeshkhind, Pune 411007, India}

\author[0000-0003-0144-4052]{Maxim Markevitch}
\affiliation{NASA Goddard Space Flight Center, Code 662, Greenbelt, MD 20771}

\author[0000-0003-0297-4493]{Paul Nulsen}
\affiliation{Center for Astrophysics $|$ Harvard \& Smithsonian, 60 Garden St.,
Cambridge, MA 02138, USA}
\affiliation{ICRAR, University of Western Australia, 35 Stirling Hwy, Crawley, WA 6009, Australia}

\author[0000-0001-6421-054X]{Christopher P.~O'Dea}
\affiliation{University of Manitoba, Dept. of Physics and Astronomy, Winnipeg, MB R3T 2N2, Canada}

\author[0000-0003-3175-2347]{Scott Randall}
\affiliation{Center for Astrophysics $|$ Harvard \& Smithsonian, 60 Garden St.,
Cambridge, MA 02138, USA}

\author[0000-0002-4864-4046]{Somak Raychaudhury}
\affiliation{Ashoka University, Dept. Of Physics, Sonipat, Haryana-131029, India}

\author[0000-0003-3295-6595]{Sravani Vaddi}
\affiliation{Arecibo Observatory, NAIC, HC3 Box 53995, Arecibo, Puerto Rico, PR 00612, USA}

\author[0000-0001-8121-0234]{Alexey Vikhlinin}
\affiliation{Center for Astrophysics $|$ Harvard \& Smithsonian, 60 Garden St.,
Cambridge, MA 02138, USA}

\author[0000-0003-3175-2347]{John Zuhone}
\affiliation{Center for Astrophysics $|$ Harvard \& Smithsonian, 60 Garden St.,
Cambridge, MA 02138, USA}

\begin{abstract} 
\noindent To examine how AGN feedback shapes the intracluster medium (ICM) and fuels black hole accretion in the cool-core galaxy cluster Abell 2597, we present deep ($\sim$600 ks) \chandra\ X-ray observations complemented by archival GMRT radio and SINFONI near-infrared data. Radio-mode AGN activity has inflated seven X-ray cavities and driven one to three potential weak shocks ($\mathcal{M} \sim 1.05-1.14$) extending to $\sim 150$ kpc, suggesting recurrent outbursts occurring on $\sim 10^7$ year timescales. We also detect a narrow, $\sim$57 kpc X-ray surface brightness deficit—a potential plasma depletion layer—likely shaped by residual sloshing motions that amplified magnetic fields and/or displaced gas within the cluster core. Although the AGN injects $\sim 10^{44}$ erg s$^{-1}$ of energy, comparable to the cluster’s cooling luminosity, radiative cooling persists at $\sim$15 \mrate, replenishing the billion solar mass cold gas reservoir at the heart of the brightest cluster galaxy. Sustaining this level of activity requires a continuous fuel supply, yet the estimated Bondi accretion power ($\sim 2 \times 10^{43}$ erg s$^{-1}$) falls an order of magnitude short of the observed cavity power, suggesting that ``hot'' gas fueling is insufficient. Instead, archival ALMA observations continue to support a chaotic cold accretion scenario, where turbulence-driven condensation fuels the AGN at rates exceeding Bondi accretion, sustaining a self-regulated feedback cycle that repeatedly shapes the core of Abell 2597. 
\\\\
\noindent\textit{Unified Astronomy Thesaurus concepts:} Galaxy clusters (584); Active galactic nuclei (16); Shocks (2086)
\end{abstract}

\section{Introduction} \label{sec:intro}
Galaxy clusters—the largest collapsed structures in the Universe—contain hundreds to thousands of galaxies embedded in hot ($\sim 10^7 - 10^8$ K), diffuse plasma known as the intracluster medium (ICM). Nearly half of all clusters exhibit a ``strong cool core,” where the central region hosts an overdensity of rapidly cooling, X-ray-bright gas  \citep{hudson_what_2010, eckert_cool-core_2011, andrade-santos_fraction_2017, rossetti_cool-core_2017}. As the gas cools, it loses pressure support, resulting in a continuous inflow of cooler, multiphase gas toward the central brightest cluster galaxy (BCG). In the classic ``cooling flow'' model, this process should fuel star formation rates of up to several hundred solar masses per year.  However, observations consistently fail to detect such extreme star formation, suggesting that the remaining cooling is hidden from view \citep[e.g.,][]{fabian_hidden_2022, fabian_hidden_2023}, and/or some mechanism, most likely mechanical feedback from the supermassive black hole (SMBH) at the center of the BCG, is heating the gas and suppressing the cooling flow (\citealt{veilleux_galactic_2005, mcnamara_heating_2007, fabian_observational_2012, mcnamara_mechanical_2012, kormendy_coevolution_2013,gaspari_linking_2020}, for reviews).

Abell 2597 ($z=0.0821$), a nearby cool-core galaxy cluster, serves as a canonical example of AGN feedback regulating cooling in cluster cores, preventing extreme star formation in BCGs and contributing to the steep decline in the galaxy luminosity function at high luminosities \citep{bower_breaking_2006, croton_many_2006}. Within the cluster's rapidly cooling hot atmosphere, the BCG's SMBH expels powerful radio jets, inflating a $\sim 30$ kpc network of buoyantly rising bubbles observed as X-ray `cavities' \citep{tremblay_multiphase_2012}. As the bubbles rise and expand, they heat the surrounding plasma \citep{churazov_cooling_2002, tremblay_multiphase_2012}, reducing the classical cooling flow rate from $\sim 500$\mrate\ to $\sim 20$–$75$\mrate, as measured by O VI ultraviolet emission tracing $\sim 7 \times 10^5$ K gas \citep{oegerle_fuse_2001}. In addition to mitigating intracluster gas cooling, these bubbles likely uplifted the $\sim 30$ kpc multiphase filamentary nebula observed in the cluster core \citep{tremblay_residual_2012, tremblay_galaxy-scale_2018}.

Recent Atacama Large Millimeter/submillimeter Array (ALMA) CO(2-1) observations of the BCG’s cold gas revealed three redshifted ($+300$ \kms) absorption lines against the SMBH’s mm-synchrotron continuum, tracing infalling clouds delivering $\sim 0.1$ to a few \msun \yr\ of gas to the black hole \citep{tremblay_cold_2016}. These observations are consistent with recent works showing that fueling SMBHs in cool-core clusters may be predominantly driven by turbulent cold clouds via chaotic cold accretion (\citealt{gaspari_chaotic_2013,gaspari_raining_2017}) rather than hot accretion flows (\citealt{bondi_spherically_1952}). Regardless of the dominant accretion phase, the inflowing gas fuels the next round of AGN activity, driving jets and bubbles that uplift cool, low-entropy gas, sustaining a long-lived, self-regulating, galaxy-spanning ``fountain” with the black hole acting as its ``pump'' \citep{tremblay_galaxy-scale_2018}.

To map the hot phase of the billion solar mass fountain, \cite{tremblay_multiphase_2012} analyzed $\sim 120$ ks of \textit{Chandra} data, revealing a highly asymmetric surface brightness distribution of the ICM and an extended 30 kpc network of X-ray cavities. The largest cavity, with a projected length of $\sim 35$ kpc, aligns with 330 MHz radio emission, and spectral maps revealed an arc of hot, high-entropy gas bordering the cavity's inner edge. Given its morphology and location, they considered two interpretations: (1) the arc may trace ambient $\sim$keV gas that appears hotter in contrast to adjacent cooler gas—a line-of-sight projection effect rather than a genuinely heated structure; or (2) it may represent gas heated as it rushed inward to refill the wake left by the buoyantly rising bubble, thermalizing the cavity’s enthalpy and depositing energy into the ICM. In the second scenario, the hot arc is a direct signature of AGN-driven heating associated with the cavity’s ascent. However, the limited depth of the X-ray data prevented a definitive conclusion.

Deep \chandra\ observations, often spanning hundreds of kiloseconds (ks) to megaseconds, have been instrumental in examining the imprints AGN feedback leaves on the hot ICM. These long exposures have led to the discovery of multiple generations of cavities \citep[e.g., Perseus, A2052;][]{fabian_deep_2003, blanton_very_2011}, weak shock fronts surrounding powerful outbursts 
\citep[e.g., Hydra A, NGC 5813, Cygnus A, MS 0735.6+7421, Abell 2052, Hercules A;][]{nulsen_cluster-scale_2005, randall_very_2015, snios_cocoon_2018, liu_agn_2019, ubertosi_cocoon_2025}, and in extreme cases, ripples--interpreted as multiple weak shock fronts--propagating outward into the large-scale ICM \citep[e.g., Perseus;][]{fabian_very_2006}, while also examining the role of hot gas in fueling AGN activity \citep{russell_radiative_2013, russell_inside_2015, bambic_agn_2023}.

Building on this legacy of deep \chandra\ observations, we present an additional 480 ks of data for Abell 2597, increasing the total exposure to 590 ks. We also incorporate archival low-frequency radio observations from the Giant Metrewave Radio Telescope (GMRT), and near-infrared integral field unit observations from the Spectrograph for INtegral Field Observations in the Near Infrared (SINFONI) on the Very Large Telescope (VLT). This paper is structured as follows: Section \ref{sec:obs_data} describes the deep X-ray data and archival radio, ultraviolet, and near-infrared observations used; Section \ref{sec:overall_icm} presents a comprehensive description of both newly identified and previously reported structures in the ICM; Sections  \ref{sec:heating_icm} and \ref{subsec:origin_channel} examine the heating and cooling balance within the cluster, fuel sources for the SMBH, and the potential origin of the features observed. 

In this work, we adopt a cosmology with $H_0$ = 70 \kms\ Mpc$^{-1}$, $\Omega_M$ = 0.27, and $\Omega_\Lambda$ = 0.73. At $z=0.0821$, 1\arcsec\ corresponds to 1.549 kpc, with angular size and luminosity distances of 319.4 Mpc and 374.0 Mpc, respectively, and a Universe age of 12.78 Gyr. Unless stated otherwise, all images are centered on the A2597 BCG nucleus at right ascension 23h 25m 19.7s, declination -12$^\circ$ 07\arcmin\ 27\arcsec\ (J2000), with east left and north up. All quoted uncertainties represent $1\sigma$ confidence intervals.

\section{Observations and Data Reduction} \label{sec:obs_data}

Below, we describe all new and archival observations of Abell 2597 used in this study and also summarize them in Table \ref{tab:observations}. To ensure transparency and reproducibility, all Python codes and Jupyter Notebooks used for data analysis are publicly available on Zenodo under a Creative
Commons Attribution license, doi:\url{10.5281/zenodo.17610478}.

\begin{deluxetable*}{cccccccccc}
\tabletypesize{\footnotesize}
\tablecaption{\textsc{Summary of Abell 2597 Observations}\label{tab:observations}}
  \tablehead{
    \colhead{Waveband/Line} &
    \colhead{Facility} &
    \colhead{Instrument/Config.} &
    \colhead{Exp. Time} &
    \colhead{Prog./Obs.ID (Date)} &
    \colhead{Reference}\\[-8pt]
    \colhead{(1)} &
    \colhead{(2)} &
    \colhead{(3)} &
    \colhead{(4)} &
    \colhead{(5)} &
    \colhead{(6)} 
}
  \startdata
X-ray (0.2 - 10 keV)& \textit{Chandra}    &  ACIS-S/VFAINT & 52.20 ks & 6934 (2006 May 1) & Tremblay et al. (2012a, 2012b) \\
\nodata   &  \nodata & \nodata  &  60.11 ks & 7329 (2006 May 4) & Tremblay et al. (2012a, 2012b) \\
\nodata    &  \nodata & \nodata  &  69.39 ks & 19596 (2017 Oct 8) & (Large Program 18800649) \\
\nodata   &  \nodata & \nodata  &  44.52 ks & 19597 (2017 Oct 16) & \nodata \\
\nodata &  \nodata & \nodata  &  14.34 ks & 19598 (2017 Aug 15) & \nodata \\
\nodata  &   \nodata & \nodata  &  24.73 ks & 20626 (2017 Aug 15) & \nodata \\
\nodata  & \nodata & \nodata  &  20.85 ks & 20627 (2017 Aug 17) & \nodata \\
\nodata &  \nodata & \nodata  &  10.92 ks & 20628 (2017 Aug 19) & \nodata \\
\nodata  &  \nodata & \nodata  &  56.36 ks & 20629 (2017 Oct 3) & \nodata \\
\nodata   &  \nodata & \nodata  &  53.4 ks & 20805 (2017 Oct 5) & \nodata \\
\nodata    & \nodata & \nodata  &  37.62 ks & 20806 (2017 Oct 7) & \nodata \\
\nodata   &  \nodata & \nodata  &  79.85 ks & 20811 (2017 Oct 21) & \nodata \\
\nodata  &  \nodata & \nodata  &  62.29 ks & 20817 (2017 Oct 19) & \nodata \\
\hline
\textit{K}-band & VLT & SINFONI + AO & 3600 s & 093.B-0638(C) (2014 Jul 19) & this paper \\
FUV & HST & COS G140L & 13952.640 s & 30-Oct 2014  & Vaddi \& Omoruyi et al (in prep) \\
Radio & GMRT & Band 4  & 85 mins & 05DAG01, Obs. 1413  & this paper \\
 &  &   &   & (20-Mar 2004) &  \\
\nodata & uGMRT & Band 3  & 32 mins & 38\_056, Obs. 12503  & this paper \\
 &  &   &   & (07-Oct 2020) &  \\
\hline
  \enddata
 \tablecomments{Summary of all new and archival observations of Abell 2597 presented in this paper. The observations are presented in descending order of wavelength, from X-ray through radio.
   (1) Waveband or Line Name targeted by the observation;
   (2) telescope used;
   (3) instrument (and aperture/detector) used for observation;
   (4) on-source exposure time;
   (5) date of observation;
   (6) reference to publication(s) where the listed data first appeared or were otherwise discussed in detail.
}
\vspace{-8mm}
\end{deluxetable*}

\vspace{-10mm}
\subsection{Chandra X-ray Observations}
\label{subsec:chandraobs}

As part of Cycle 18 program 18800649 (PI: Tremblay), the \chandra\ X-ray Observatory observed Abell 2597 for a total of 474.3 ks in August and October 2014 (ObsIDs 19596, 19597, 19598, 20626, 20627, 20628, 20629, 20805, 20806, 20811, and 20817). We also included 112.3 ks from two older, previously published observations (ObsIDs 6934 and  7329, PI: Clarke; \citealt{mcnamara_discovery_2001, clarke_low-frequency_2005,tremblay_multiphase_2012, tremblay_residual_2012}), leading to a total of 586.6 ks across all fourteen ACIS-S observations. The datasets used in this study are archived in the Chandra Data Collection\dataset[doi:10.25574/cdc.384]{https://doi.org/10.25574/cdc.384}. 

The ObsIDs were processed using the latest versions of \textsc{ciao} v4.16 (\chandra\ Interactive Analysis of Observations; \citealt{fruscione_ciao_2006}) and \textsc{CALDB} v4.11.5. Data were reprocessed with \textit{chandra\_repro}, and flares were identified and filtered using the ChIPS routine \textsc{lc\_clean}. Few significant flares were detected, leading us to reject a minimal 2.76 ks. Point sources were identified using a wavelet decomposition technique \citep{vikhlinin_catalog_1998}, inspected, and masked. Although A2597 remains a steady source of X-rays, the ACIS optical blocking filter has degraded over time, due to significant contaminant buildup. Consequently, the earlier \chandra\ observations used have count rates nearly twice as high as those obtained a decade later. Across the total $\sim 600$ ks exposure, the observations contain 2.1 million counts within $r_{500} =  592\farcs8$.

\subsubsection{Imaging} 
\label{subsubsec:chandra_imaging}

\begin{figure*}
\includegraphics[width=\linewidth]{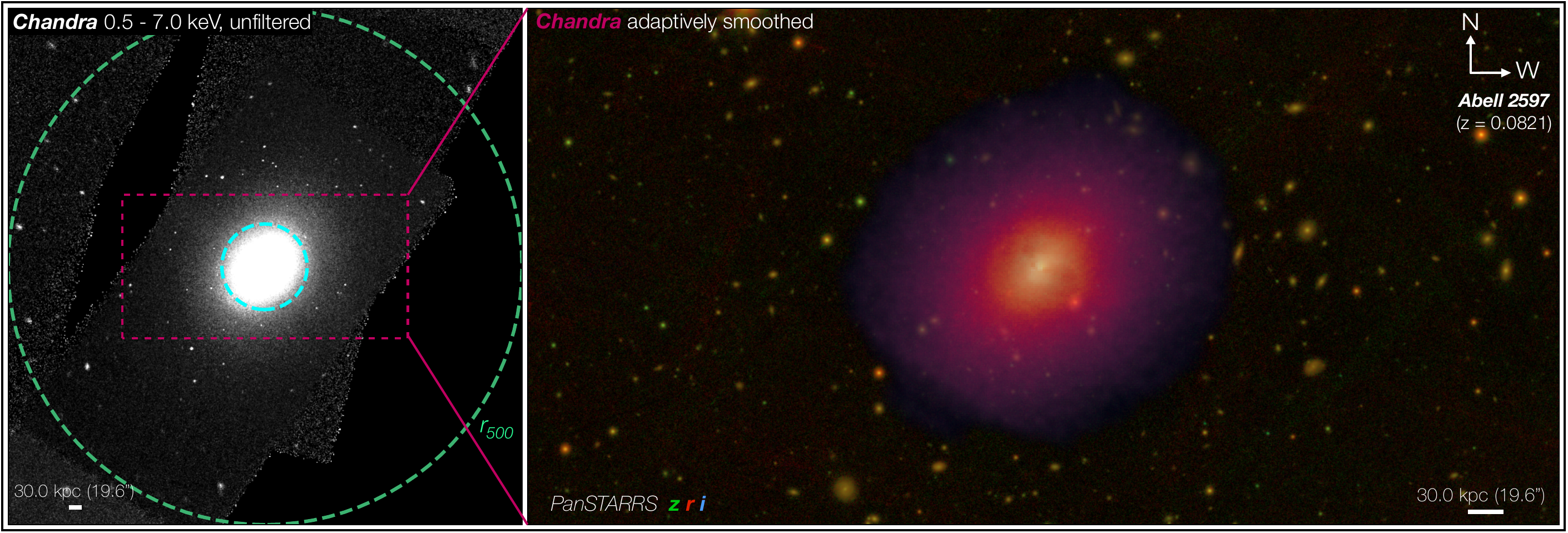}
\vspace{-5mm}
\caption{\textit{Left:} Merged, background-subtracted, and exposure-corrected 0.5--7 keV \chandra\ image of Abell 2597, shown out to $r_{500} = 592\farcs8$ (green dashed circle). The central 150 kpc is marked by a cyan dashed circle, and the magenta dashed box outlines the region shown in the right panel. \textit{Right:} PanSTARRS \textit{z, r, i} image of the central $\sim 500 \times 250$ kpc of the cluster overlaid with an adaptively smoothed version of the merged \chandra\ observations.}
\label{fig:chandra_obs_flux}
\end{figure*}

To create X-ray images of the ICM, we merged the ObsIDs with the \textsc{merge\_obs} \texttt{CIAO} script. Figure \ref{fig:chandra_obs_flux} presents the resulting merged observations out to $r_{500}$ (green dashed circle) at \textit{Chandra's} native 0\farcs5 resolution, corresponding to a physical scale of $\sim 0.75$ kpc.

To enhance the visibility of subtle features, we constructed an adaptively smoothed surface brightness map using a variable-width Gaussian kernel, requiring that each smoothed region contained at least 100 counts. The resulting X-ray image is overlaid on a false-color ($z, r, i$) PanSTARRS optical image of the cluster’s central 500 × 250 kpc region in the right panel of Figure \ref{fig:chandra_obs_flux}. 

To improve the visibility of structures within the central $180$ kpc of the cluster (dashed cyan circle), we applied a Gaussian Gradient Magnitude (GGM) filter and an adaptive GGM filter to the point-source-removed, merged image. These filters were implemented using the publicly available code described in \citet{sanders_detecting_2016, sanders_studying_2022}, and were applied at multiple spatial scales to highlight features across a range of sizes; the resulting maps are shown in Figures \ref{fig:xray_features} and \ref{fig:xray_radio}. Separately, we also created an unsharp-masked image by subtracting a $9\farcs8$-Gaussian smoothed version of the image (corresponding to $\sim$10 ACIS pixels) from one smoothed by $0\farcs98$ (corresponding to $\sim$2 ACIS pixels) \citep[e.g.,][]{hlavacek-larrondo_extreme_2012}, which is used to enhance surface brightness depressions in selected figures shown in Section \ref{sec:overall_icm}.

\subsubsection{Spectral Profiles} 
\label{subsubsec:chandra_spectral_prof}

To derive the large-scale spectral properties of A2597's ICM (e.g. $M_{500}$, $R_{500}$), we followed the methodology presented in \cite{vikhlinin_chandra_2006}. A full description of this process is provided in Section 2.3 of \cite{omoruyi_beads---string_2024}, where it was applied to SDSS J1531+3414. 

In brief, we extracted spectra from concentric annuli centered on the X-ray peak ($\mathrm{RA} = 23^{\mathrm h}25^{\mathrm m}25^{\mathrm s}.19$, $\mathrm{Dec} = -12^\circ07'27\farcs4$ (J2000)) from the central 8\farcs5 to $r_{500}$. Background estimates were obtained from each ObsID’s corresponding blank-sky event file, using the same region as the source. The profile was modeled with the modified $\beta$-model from \citet{vikhlinin_chandra_2006} (Equation 3). From the best-fit model, we derived the three-dimensional gas density $\rho_g(r)$ to compute the total gas mass $M_g$ and the classical cooling rate $\dot{M}_{\text{cool}}$.

A large-scale temperature profile was obtained using 12 logarithmically spaced annuli extending to $r_{500}$, with each annulus containing at least 74,000 net counts after background subtraction. The temperature was modeled using the three-dimensional analytic form from Equation 6 in \citet{vikhlinin_chandra_2006} and then projected. Using the derived temperature and gas density profiles, we calculated the total cluster mass within radius $r$ under the assumption of hydrostatic equilibrium \citep{sarazin_x-ray_1988}, and adopted $M_{500}$ and $R_{500}$ from the $Y_X-M$ scaling relation \citep{vikhlinin_chandra_2009}.

To better examine features in the cluster core, we extracted spectra from 12 linearly spaced concentric annuli extending to $\sim 150$ kpc, marked by the cyan circle in Figure \ref{fig:chandra_obs_flux}. For each ObsID, spectra were extracted in the 0.5–7 keV range, with background subtraction performed using blank-sky files. 

The spectra were fit with a \textsc{phabs*apec} model in \textsc{XSPEC}, fixing $N_H = 2.252 \times 10^{20}$ cm$^{-2}$ \citep{bekhti_hi4pi_2016} and $Z = 0.3 Z_\odot$ \citep[e.g.,][]{panagoulia_volume-limited_2014}, while allowing \textit{kT} and the normalization $N(r)$ to vary. The fitted values were used to derive the electron density profile, assuming spherical symmetry, from which the pressure $P\equiv 1.83n_ekT$ and entropy $K \equiv kTn_e^{-2/3}$ were computed. For the abundance profile, we adopted the same fitting procedure as above. However, to ensure sufficient signal-to-noise in the outer regions, we used the 12 logarithmically spaced annuli described earlier and allowed the abundance to vary freely.

\subsubsection{Spectral Maps} 
\label{subsubsec:chandra_spectral_maps}

To generate high-resolution temperature, pseudo-pressure, and pseudo-entropy maps, we applied the Contour Binning algorithm \citep{sanders_contour_2006}, which groups pixels along surface brightness contours until a specified signal-to-noise threshold is met. We first created an adaptively smoothed, point-source-free image with S/N = 15 in the 0.5–7 keV band and defined bins using a geometric constraint factor of C = 2 to maintain compact shapes. Bin sizes were set to achieve 5–10\% fractional uncertainties in the fitted temperature values, based on the $1\sigma$ \textsc{XSPEC} confidence intervals. Within the central 60 kpc, this accuracy was reached using a spectral S/N of 50 ($\sim 2500$ net counts per bin), while a higher S/N of 80 ($\sim 6400$ net counts per bin) was used for bins at larger radii to maintain reliable fits in lower surface brightness regions. We then used J. Sanders’ \texttt{make\_region\_files} to create \textsc{CIAO}-compatible region files, extracting spectra for each bin across all ObsIDs and estimating the background from the corresponding blank-sky files. The summed spectra were fit simultaneously in \textsc{XSPEC} using an absorbed MEKAL model, with redshift, Galactic $N_H$, and metallicity fixed to the same values used in the spectral profile fitting.

Using the best-fit temperature $kT$ and normalization $N(r)$, we constructed projected temperature, pseudo-entropy, and pseudo-pressure maps with J. Sanders’ \texttt{paint\_output\_images}. The temperature map was derived directly from $kT$, while pseudo-entropy and pseudo-pressure maps assumed the electron density scales as $\sqrt{N(r)}$, yielding units of keV cm$^{-5/2}$ arcsec$^{-1}$ for pressure and keV cm$^{5/3}$ arcsec$^{2/3}$ for entropy. For the abundance map, we followed the same procedure but increased the minimum S/N per bin to 100 and used circular bins to ensure robust fits. All spectral maps shown in Section \ref{subsec:spectral_maps} only include bins where the fractional error is smaller than 10\%, except for the abundance map, where we show fractional errors $< 20\%$. 

\subsection{Archival GMRT Radio Observations}

Abell 2597 was observed multiple times with the GMRT \citep{swarup_giant_1991}, and the upgraded GMRT (uGMRT; \citealt{gupta_upgraded_2017, reddy_wideband_2017}) between 2000 and 2020 at frequencies ranging from 150 - 630 MHz. However, most legacy GMRT datasets were severely affected by radio frequency interference (RFI), rendering them unsuitable for spatially resolved imaging. Only two datasets were successfully recovered for use in this work: a legacy GMRT 610 MHz observation (Project ID:05DAG01, PI: David Alan Green), and a uGMRT Band 3 observation centered at 400 MHz (Project ID: 38\_056, PI: Nithyanandan Thyagarajan). 

The 610 MHz legacy data, obtained on 2004 March 20 (Observation No. 1413), was reduced using the CASA Pipeline-cum-Toolkit for Upgraded GMRT data Reduction (CAPTURE) pipeline \citep{kale_capture_2021} with $\tt{CASA}$ version 6.1.0.118. The raw data were converted from the Long-Term Archive (LTA) format to FITS using the $\tt{LISTSCAN}$ and $\tt{GVFITS}$ utilities provided by the CAPTURE package and loaded into $\tt{CASA}$. The pipeline then carried out the standard amplitude, phase and bandpass calibration procedures using bright and nearby calibrators. Additional editing of bad data was carried out using the tasks $\tt{TFCROP}$, $\tt{RFLAG}$, and the $\tt{EXTEND}$ mode of $\tt{FLAGDATA}$ in $\tt{CASA}$. Finally, we performed two rounds of phase-only self-calibration and three rounds of amplitude and phase self-calibration on the data. The final 610 MHz image has $\sigma_{\rm rms}\sim$9.3 mJy beam$^{-1}$ and a synthesized beam size of $11\farcs43 \times 4\farcs87$.

The 400 MHz uGMRT Band 3 observations were carried out on 2020 October 7 (Observation no. 12503), during which A2597 (also cataloged as 2325-121) was used as a phase calibrator. The observations were taken with the upgraded GMRT system, which offers significantly broader bandwidth and improved sensitivity compared to the legacy GMRT. Although Abell 2597 was not the primary science target, the data captured extended low-frequency emission with greater spatial fidelity than the 610 MHz legacy observations. The image quality is limited primarily by dynamic range rather than RFI due to the high intrinsic flux density of A2597 ($\sim$7 Jy at 400 MHz). Calibration and imaging were carried out using the SPAM pipeline \citep{intema_gmrt_2017, intema_ionospheric_2009}, which applies automated direction-dependent corrections. The full bandwidth was divided into six subbands, each independently processed through RFI flagging, bandpass, flux, and phase calibration using models from \citealt{scaife_broad-band_2012}. The calibrated subbands were then jointly deconvolved with WSClean \citep{offringa_wsclean_2014} using multiscale cleaning and Briggs weighting (robust = $-1.5$) to produce the final continuum image with a synthesized beam of $4\farcs5 \times 4\farcs5$ at a position angle of $0\degr$. The resulting images are presented in Section \ref{subsec:radio_xray}.

\begin{figure*}
\centering
\includegraphics[width=0.95\linewidth]{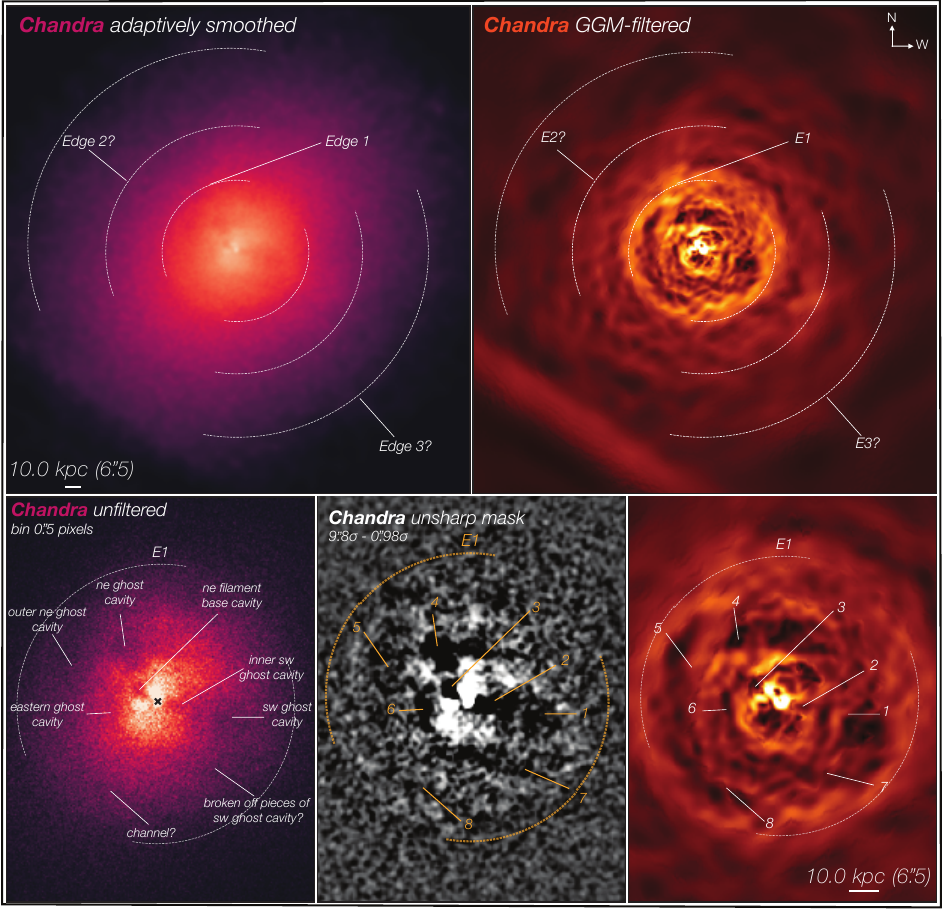}
\caption{Features in the ICM of Abell 2597 identified by eye. \textit{Top:} Adaptively smoothed (left) and GGM filtered (right) images of the central $\sim 300 \times 180$ kpc cluster environment, highlighting the approximate locations of three surface brightness edges (dashed white arcs). The image scaling was selected to maximize the visibility of all three edges simultaneously. \textit{Bottom:} Central $\sim 75$ kpc region showing an unfiltered 0\farcs5 pixel binned \textit{Chandra} image (left), unsharp mask image (middle) and a GGM-filtered image (right). Seven cavities and an X-ray channel are identified and labeled. Note that in the top right GGM filtered image, the diagonal line in the lower left corner is an artifact of the present chip gap.}
\label{fig:xray_features}
\end{figure*}

\subsection{Archival SINFONI Observations}

In 2014, the Spectrograph for INtegral Field Observations in the Near Infrared (SINFONI; \cite{eisenhauer_sinfoni_2003}) on the Very Large Telescope observed the nucleus of A2597. Under ESO Program ID 093.B-0638(C) (PI: Raymond), the near-infrared observations were performed in the K-band filter (1.95–2.45 \micron), providing integral field data cubes covering the $H_2$ (ro)vibrational lines. The observations were performed using the Laser Guide Star Adaptive Optics (AO) mode \citep{bonnet_implementation_2003}. Strap-length limits required that tip/tilt correction, which compensates for small-scale variations in the wavefront caused by atmospheric turbulence and improves spatial resolution, be attempted on the A2597 nucleus. The correction, however, was unsuccessful, for the nucleus lacked sufficient contrast against the blurred galaxy caused by atmospheric conditions. Consequently, the observations were conducted in a ``seeing enhancer mode" which enables us to achieve a spatial resolution of 0\farcs6 (FWHM), only a moderate increase in spatial resolution relative to the previous 1\arcsec seeing-limited images of the nucleus published in \cite{oonk_distribution_2010}. 

The SINFONI-AO data were reduced using \textsc{esorex} (3.13.6) workflow 3.3.2 in the EsoReflex environment \citep{freudling_automated_2013}. Emission lines in the resulting data cube were modeled using the spectroscopic analysis toolkit \textsc{pyspeckit} \citep{ginsburg_pyspeckit_2022}. Continuum emission from the host stellar component was subtracted using nearby line-free regions, and the H$_2$ emission lines were fit with single-component Gaussian profiles.
\
\section{The Deepest View of the Hot ICM} 
\label{sec:overall_icm}
\subsection{X-ray Morphology and Features}
\label{subsec:xray_features}
Figure \ref{fig:xray_features} presents background-subtracted, exposure-corrected \textit{Chandra} X-ray images of Abell 2597 in the 0.5–7.0 keV band. The top panel shows adaptively smoothed (left) and GGM-filtered (right) X-ray images of the cluster’s central $\sim 300$ kpc $\times$ 180 kpc environment. The X-ray emission is centrally peaked, characteristic of cool-core clusters, and gradually fades into a diffuse halo. The overall morphology is broadly asymmetric, with prominent extensions to the northeast and southwest. Additionally, three low-contrast, nearly elliptical surface brightness edges are present, with their approximate locations marked by dashed white arcs. Although Edge 1 is clearly visible in the images, Edges 2 and 3 are considerably more subtle; without the overlaid arcs, the latter two features would be difficult to distinguish by eye. The smoothing and scaling used in the large-scale images were selected to maximize the visibility of all three edges simultaneously.

The most prominent structures lie within the innermost surface brightness edge. To highlight these features, the lower three panels of Figure \ref{fig:xray_features} zoom in on the central $\sim 56$ kpc (37\farcs0). The X-ray peak, marked by a cross in the left panel, coincides with the optical center of the BCG \citep{tremblay_multiphase_2012}. Around the peak, the emission forms a backward ‘S’ shape due to surface brightness deficits to the northeast and southwest. This feature is more clearly visible in the unsharp-masked and GGM-filtered maps to the right. The deficits are labeled numerically, with a full list provided in Table \ref{tab:xray_features}.

The first 32 ks \chandra\ X-ray observations of Abell 2597 by \cite{mcnamara_discovery_2001} (hereafter \citetalias{mcnamara_discovery_2001}) identified two prominent cavities in the ICM: one to the northeast (Feature 4) and one to the southwest (Feature 1). Their distance from the central AGN and lack of associated radio emission led to their classification as “ghost” cavities, indicative of an older epoch of AGN feedback. However, subsequent analysis by \cite{clarke_low-frequency_2005} (hereafter \citetalias{clarke_low-frequency_2005}) using residual imaging revealed that the southwestern cavity was spatially connected to an inner cavity (Feature 2). This suggested a possible link between the outer cavity and the central radio source, though it remained unclear whether these structures were distinct cavities or part of a singular bubble.

Deeper \chandra\ observations, with nearly four times the prior exposure time (120 ks), presented in \cite{tremblay_multiphase_2012, tremblay_residual_2012} (hereafter T12a  and T12b) discovered that the network of X-ray cavities was more extended than previously thought. Notably, the western ghost cavity described by \citetalias{mcnamara_discovery_2001} and \citetalias{clarke_low-frequency_2005} was now more clearly one large, teardrop-shaped cavity extending $\sim 35$ kpc in projected length, referred to here as the southwestern cavity (Features 1 and 2). \citetalias{tremblay_residual_2012} also identified an outer northeastern ghost cavity  (Feature 5).

With our additional 480 ks of data, we find that the southwestern cavity appears connected to subtle surface brightness deficits directly below it (Feature 7). Their elongated structures, however, suggest that they may not be distinct cavities but rather fragments of the original bubble encompassing Features 1 and 2, which we discuss further in Section \ref{subsubsec:cav_origin}. We also identify a potential new cavity east of the X-ray peak (Feature 6), roughly opposite the southwestern cavity. Whether it extends northeastward toward the previously identified northeastern ghost cavity (Feature 4) remains uncertain, leaving open the possibility that it shares an elongated morphology similar to the southwestern cavity. Lastly, we identify a thin, elongated surface brightness depression to the south (Feature 8). Given its morphology, we tentatively classify it as an X-ray brightness ``channel,” similar to structures observed in Abell 520 and Abell 2142 \citep{wang_merging_2016, wang_deep_2018}.
{%
\setlength{\tabcolsep}{12pt}
\begin{table}
  \caption{Notable Features in Abell 2597's ICM}
  \hspace{-8mm}
  \begin{tabular}{c c c}
    \hline\hline
    Label & Feature & Ref. \\
    \hline
    E1 & Edge 1 & \textit{New} \\
    E2? & Edge 2? & \textit{New} \\
    E3? & Edge 3? & \textit{New} \\
    1 & SW Ghost Cavity & \citetalias{mcnamara_discovery_2001} \\
    2 & Inner SW Ghost Cavity & \citetalias{clarke_low-frequency_2005} \\
    3 & NE Filament Base Cavity & \citetalias{clarke_low-frequency_2005} \\
    4 & NE Ghost Cavity & \citetalias{mcnamara_discovery_2001} \\
    5 & Outer NE Ghost Cavity & \citetalias{tremblay_multiphase_2012}, \citetalias{tremblay_residual_2012} \\
    6 & Eastern Ghost Cavity  & \textit{New} \\
    7 & Broken off Pieces of & \textit{New} \\
 & SW Ghost Cavity? &  \\
    8 & Channel & \textit{New} \\
    \hline
    \vspace{-8mm}
  \end{tabular}
  \label{tab:xray_features}
\end{table}
}

\begin{figure*}
\includegraphics[width=\linewidth]{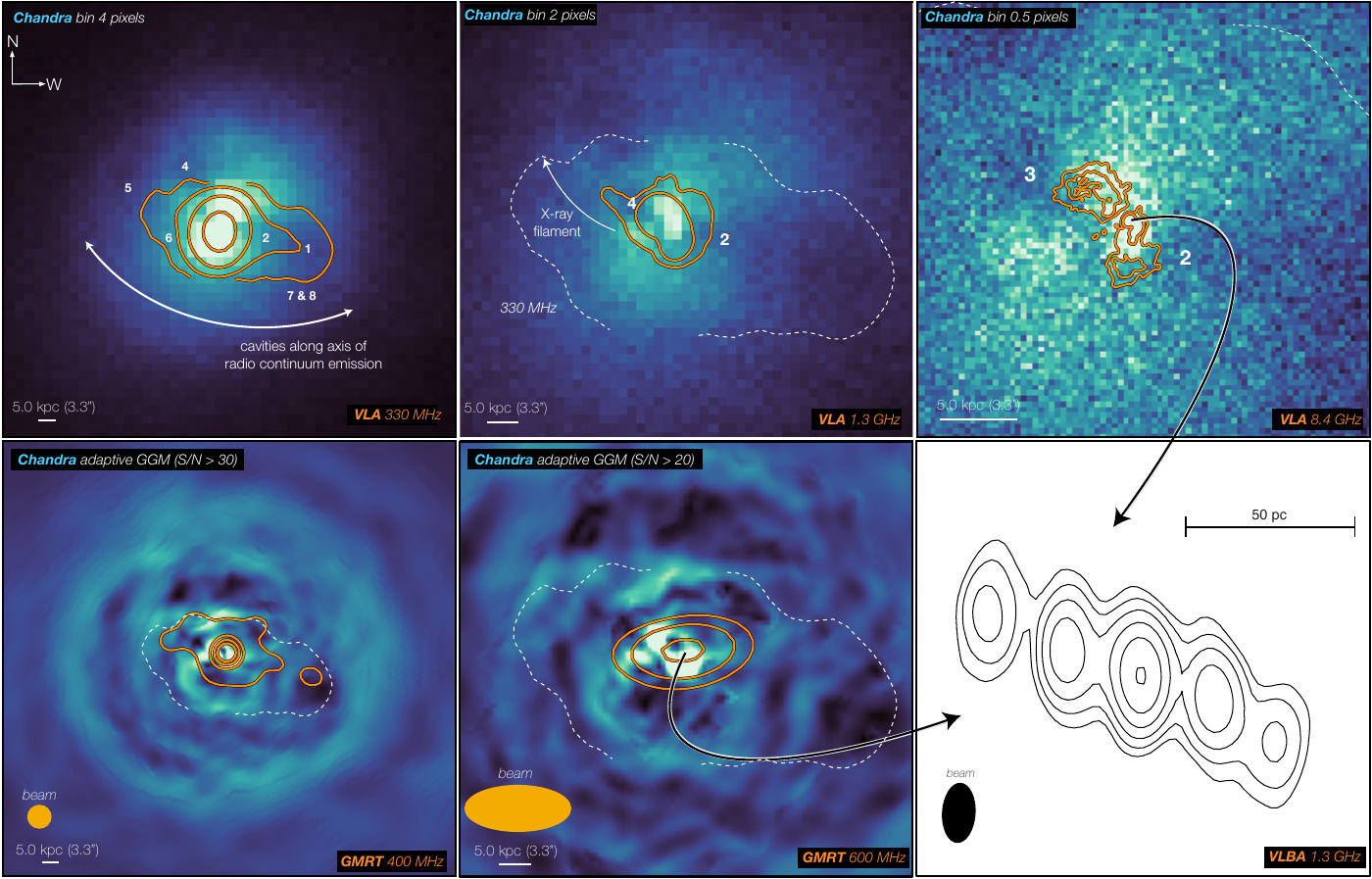}
\caption{X-ray (blue) and radio (orange contours) comparison within the cluster's central 75 kpc.\textit{Top:} X-ray images binned at 4\farcs0, 2\farcs0, and 0\farcs5 pixels, overlaid with VLA radio contours at 330 MHz, 1.3 GHz, and 8.4 GHz.  \textit{Bottom:} Adaptive GGM-filtered X-ray images with S/N$\sim 20$ (left) and S/N$\sim 30$ (center), overlaid with new GMRT 400 MHz (left) and 600 MHz (center) contours. The rightmost panel shows a zoomed-in $\sim 70$ pc view of the radio core, traced by archival 1.3 GHz VLBA continuum data, revealing the jet’s orientation at launch. All cavities align with the axes of the radio continuum emission. }
\label{fig:xray_radio}
\end{figure*}

The extensive AGN feedback activity in the cluster core may be linked to the large-scale edges seen in the top panel of Figure \ref{fig:xray_features}.  If real, these edges could represent broad, large-scale ``ripples'' in the ICM, i.e., shock waves that have weakened over time and evolved into sound waves propagating through the gas. Large-scale shocks are typically observed only in clusters with deep X-ray exposures, either as multiple weak shock fronts propagating outward into the ICM \citep[e.g., Perseus, Abell 2052;][]{fabian_deep_2003, fabian_very_2006, blanton_shocks_2009} or as cocoon shocks driven by high-pressure plasma from AGN jets inflating the radio lobes and compressing the surrounding gas, e.g., Hydra A \citep{nulsen_cluster-scale_2005}, Cygnus A \citep{snios_cocoon_2018}, and RBS 797 \citep{ubertosi_multiple_2023}. In both cases, the central AGN is the expected driver of these shock fronts, with cocoon shocks marking more recent activity from supersonic lobe expansion, while large-scale ripples are interpreted as older shocks that have traveled farther through the ICM over time.

\subsection{Jet-ICM Interaction}
\label{subsec:radio_xray}

Abell 2597 is not only home to an extensive network of X-ray cavities but also a complex system of radio sources spanning multiple frequencies. Figure \ref{fig:xray_radio} presents a multi-frequency comparison of the radio continuum and X-ray emission in Abell 2597. 

The top panel overlays VLA radio contours at 330 MHz, 1.3 GHz, and 8.4 GHz onto X-ray images binned at 4\farcs0, 2\farcs0, and 0\farcs5 pixels, illustrating how high-frequency emission from recent jet activity and low-frequency emission from older plasma correlate with X-ray structures. \citetalias{clarke_low-frequency_2005} found that the 330 MHz radio emission, which traces aged, extended plasma, is cospatial with the entire southwestern cavity. Notably, the edges of the 330 MHz source lie just inside the innermost surface brightness edge (E1). This suggests that E1 may have originated as a cocoon shock driven by the expanding radio lobes, which we explore further in Section \ref{subsec:agn_history}. 

\begin{figure*}
\includegraphics[width=\linewidth]{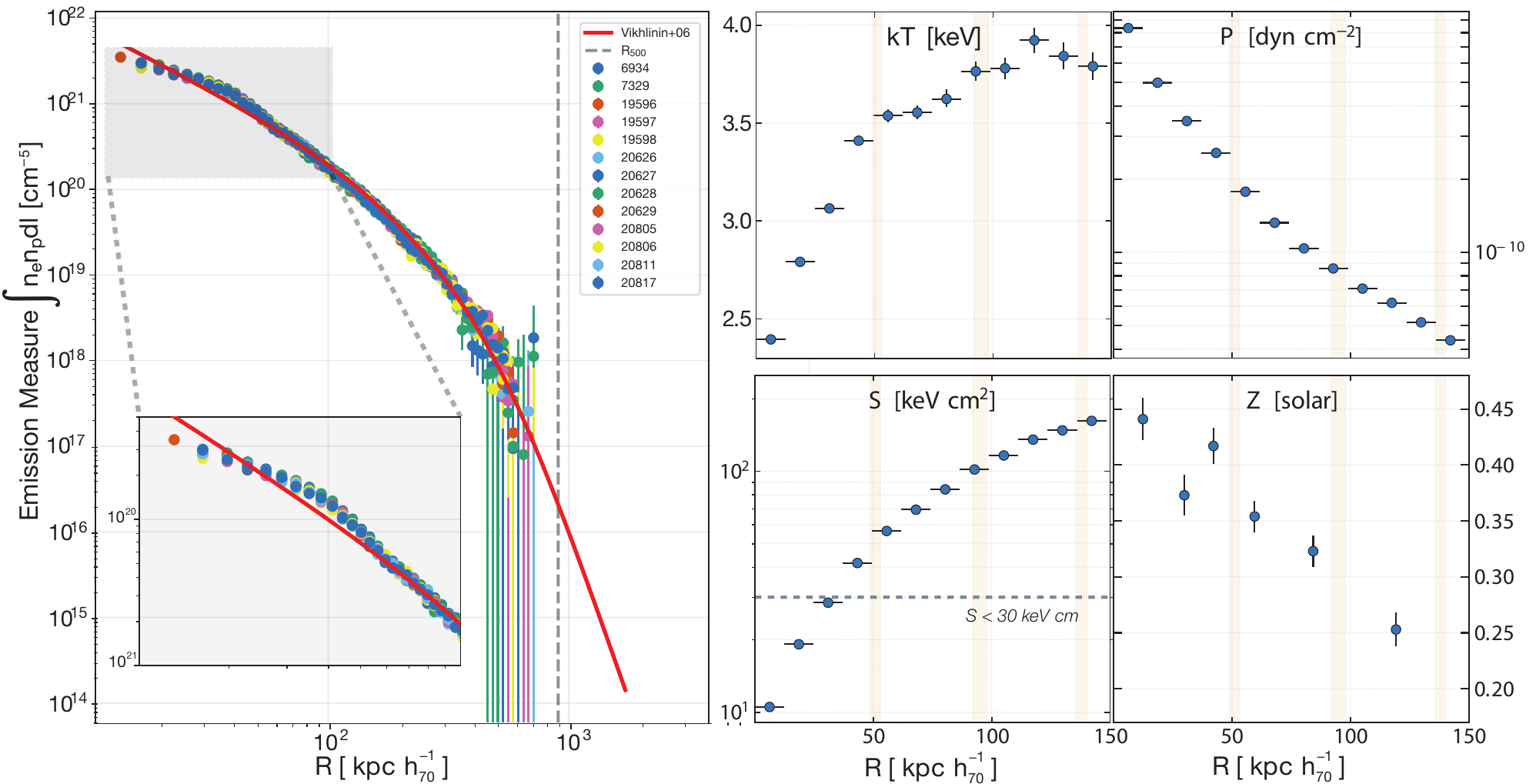}
\vspace{-5mm}
\caption{ \textit{Left: }X-ray emissivity per unit area for all \chandra\ ObsIDs (colorful points), fit with a modified beta model (red line) from \cite{vikhlinin_chandra_2006}. \textit{Right:}  Projected radial profiles of temperature ($kT$), pressure ($P$), entropy ($S$), and abundance ($Z$) within the central 150 kpc. The yellow shaded regions mark the approximate radial extent of X-ray surface brightness edges identified in Figure \ref{fig:xray_features}, extending from the inner radius of the minor axis to the outer radius of the major axis.}
\label{fig:chandra_profiles}
\end{figure*}

While the older radio plasma aligns with the outer cavities, the younger plasma primarily traces the inner cavities. The 1.3 GHz and 8.4 GHz radio sources are directly connected to the northeastern filament base cavity and the inner southwestern ghost cavity. Additionally, the large-scale 1.3 GHz emission coincides with a bright X-ray filament between the northern and eastern cavities. As noted by \citetalias{tremblay_multiphase_2012}, this alignment suggests that past jet activity may have uplifted cooler, denser gas along this direction.

The lower left panels of Figure \ref{fig:xray_radio} overlay archival GMRT 400 MHz and 600 MHz radio contours on adaptively smoothed, GGM-filtered X-ray images. The 400 MHz emission closely follows the 330 MHz source but is more compact, except for the northeastern extension, where it overlaps more with the northeastern ghost cavity than the 330 MHz emission, suggesting a common origin. The 600 MHz emission is similarly contained within the 330 MHz source but does not fully encompass it, instead extending toward the northern cavity and the base of the southwestern teardrop cavity. 

The lower right panel of Figure \ref{fig:xray_radio} provides a high-resolution VLBA 1.3 GHz view of the central 70 pc \citep{tremblay_cold_2016}, revealing the jet’s orientation at launch. While the 8.4 GHz emission traces younger plasma, the VLBA data probe smaller scales, capturing the jet’s initial trajectory. Across all frequencies, the jet direction remains largely consistent, oriented northeast to southwest, with all identified cavities closely aligning with the axis of the radio continuum emission.

\subsection{Thermodynamic Profiles of the ICM}
\label{subsec:therm_profiles}
We present thermodynamic profiles of the ICM in Figure \ref{fig:chandra_profiles}. Key cluster properties, namely the total mass $M_\Delta$ and radius $R_\Delta$, and central temperature $kT_0$, entropy $K_0$, and pressure $P_0$, are listed in Table \ref{tab:chandra_profiles}. 

The projected emission measure profile (left panel of Figure \ref{fig:chandra_profiles}) confirms the central excess characteristic of cool-core clusters. The modified $\beta$-model from \cite{vikhlinin_chandra_2006} (red line) provides a good overall fit, though deviations appear within the central 100 kpc (grey-shaded region), as shown in the inset panel, where surface brightness fluctuations coincide with the location of the X-ray cavities and inner surface brightness edge (feature E1). 

At low redshift, cool-core clusters typically exhibit central cusps in their X-ray surface brightness distributions. \cite{vikhlinin_lack_2007} characterized these cusps using the power-law index of the gas density profile, $\alpha = -2 d \log \rho_g / d \log r$, evaluated at $r = 0.04 R_{500}$. Clusters with strong cool cores generally have $\alpha > 0.7$. While \cite{vikhlinin_lack_2007} previously reported $\alpha = 1.1$ for A2597, our deeper data reveal a steeper value of $\alpha = 2.4$, further confirming the presence of a strong cool core.

The right panels of Figure \ref{fig:chandra_profiles} show projected thermodynamic profiles within the central 150 kpc, encompassing all the prominent features identified in Figure \ref{fig:xray_features}. The temperature (top left) declines toward the core, from 3.5 keV at $r \sim 100$ kpc to 2.25 keV at the center. Small fluctuations within $2$–$3\sigma$ are observed near the surface brightness edges, whose approximate radial extents are marked by yellow shaded regions. These are not interpreted here as physical discontinuities; we quantify temperature jumps more rigorously in Section \ref{subsec:cocoon}. 

{%
\setlength{\tabcolsep}{17pt}
\begin{deluxetable}{lrcc}
\tablecaption{\textsc{Thermodynamic Properties}
\label{tab:chandra_profiles}} 
\tablecolumns{4}
\tablewidth{\linewidth}
\tablehead{
\colhead{Property       
        } &
\colhead{Value                   
  }
}
\startdata
\multicolumn{2}{c}{\textit{Logarithmic Annuli}} \\
$R_{0}$  & $12.4$ kpc \\
$M_{500} - Y_X$  & $2.28 \pm 0.018$ $\times 10^{14} M_\odot$ \\
$R_{500}$ & $904.8 \pm 2.4$ kpc \\
$Z_0$  & $0.44 \pm 0.019$ $Z_\odot$\\[1ex]
\hline 
\multicolumn{2}{c}{\textit{Linear Annuli}} \\
$R_{0}$ & $6.8$ kpc \\
$kT_{0}$ & $2.40\pm 0.018$ keV \\
$K_{0}$ & $10.5\pm 0.09$ keV cm$^{2}$ \\
$P_{0}$ & $8.35 \pm 0.074$ $\times 10^{-10}$ dyn cm$^{-2}$ \\
\enddata
\tablecomments{X-ray derived thermodynamic properties of Abell 2597. Values obtained from logarithmically spaced annuli include $M_{500}$ and $R_{500}$, calculated assuming hydrostatic equilibrium and the $Y_X$--$M$ scaling relation from \cite{vikhlinin_chandra_2009}, as well as the central abundance $Z_0$. The central temperature $kT_0$, entropy $K_0$, and pressure $P_0$ were measured using linearly spaced annuli. In both sets of annuli, $R_0$ refers to the radial center of the innermost bin.}
\vspace{-12mm}
\end{deluxetable}
}
\begin{figure*}
\centering
\vspace{-10mm}
\includegraphics[width=\linewidth]{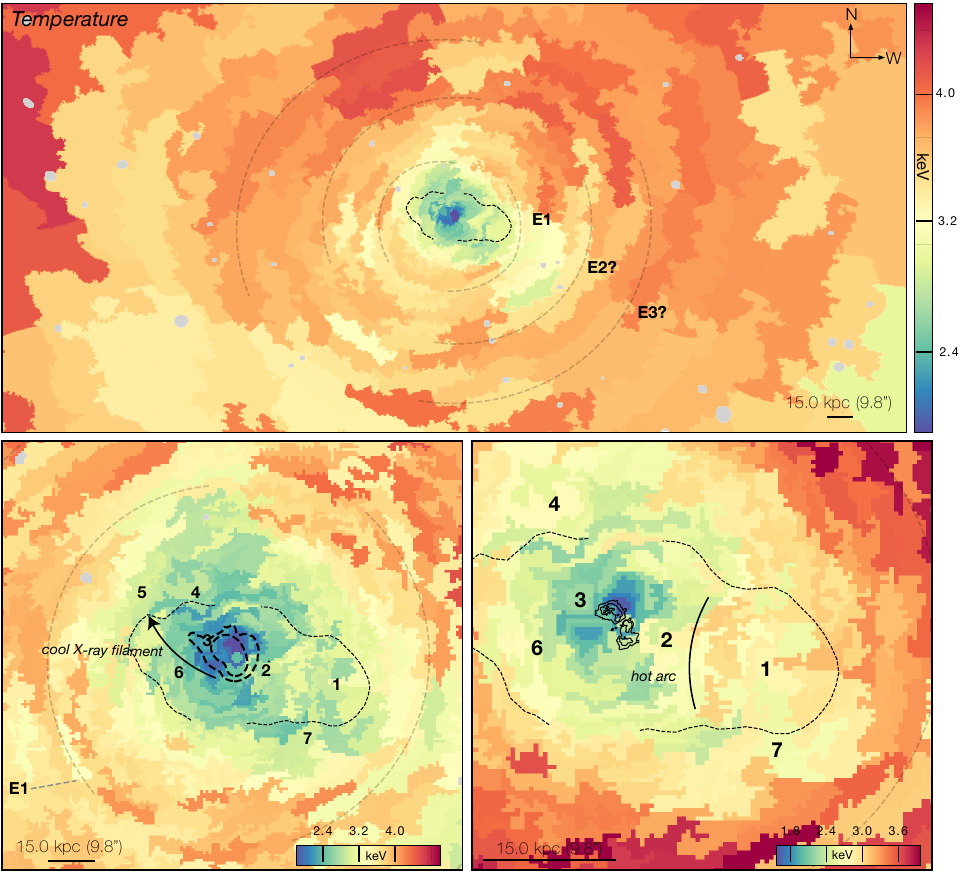}
\caption{Temperature distribution of the ICM. \textit{Top:}  Large-scale (
$\sim 620$ kpc $\times$ 320 kpc) map revealing a cool ($T< 3$ keV) core, with temperature increasing outward. Subtle temperature fluctuations are observed near the dashed elliptical arcs, which correspond to the surface brightness edges identified by eye in Figure \ref{fig:xray_features}; these may trace weak shocks from earlier AGN activity or cold fronts, as discussed in Section \ref{subsec:cocoon}. \textit{ Bottom left:} Zoomed-in view of the central 75 kpc, where the coolest gas generally traces the X-ray surface brightness distribution, including the filament identified in Figure \ref{fig:xray_radio}. \textit{Bottom right: } Hot gas borders Cavities 1 and 2, forming an arc-like structure consistent with the ``hot arc'' previously identified by \citetalias{tremblay_multiphase_2012}. The feature also appears to extend nearly fully around the rim of Cavity 1, though it blends more with the surrounding medium.}
\label{fig:xray_tmap}
\end{figure*}

The pressure profile (upper right panel) follows the expected steep radial decline for a relaxed cluster. The entropy profile (lower left panel) falls well below the 30 keV cm$^2$ threshold for star formation in cool-core clusters (dashed line) \citep{voit_conduction_2008} within the central $\sim 30$ kpc.  At the surface brightness edges, the pressure and entropy profiles are generally smooth.  The abundance profile (lower right panel) peaks at 0.44\zsun\ at the very center of the cluster and gradually declines to $\sim 0.25$ \zsun\ at 150 kpc. This central value is slightly lower than the 0.5\zsun\ metallicity of the multiphase gas measured through deep optical spectroscopy \citep{voit_deep_1997}.

\subsection{Thermodynamic Structure of the ICM}
\label{subsec:spectral_maps}

Figure \ref{fig:xray_tmap} shows the ICM temperature distribution, with the top panel displaying the large-scale structure and the lower panels focusing on the central 75 kpc. On large scales, the gas follows the expected cool-core profile, with the coolest ($T < 3$ keV) gas concentrated at the center and temperature increasing outward. Similar to the surface brightness distribution, the temperature is anisotropic, with cool gas extending along the axes of the 330 MHz radio source (black dashed contours) and to the north. Subtle temperature fluctuations are also observed near the locations of the surface brightness edges (marked by dashed elliptical arcs in Figure \ref{fig:xray_features}), particularly to the northwest and southeast. These variations may correspond to weak shocks from earlier AGN activity, but we defer quantitative analysis to the profile-based diagnostics presented in Section \ref{subsec:cocoon}.

In the central 75 kpc (lower left panel), the coolest (T $< 2.3$ keV) gas near the 8.4 GHz radio source follows a backward ‘S’ shape, mirroring the X-ray surface brightness distribution. The northern extension of cool gas aligns with the $\sim 15$ kpc soft X-ray filament identified by \citetalias{tremblay_residual_2012}, which our deeper data show extends even farther northeast, reaching $\sim 18$ kpc. 

The lower right panel provides a closer look at the spectral features of the cavities in the central 50 kpc. The large southwestern cavity (Features 1 and 2) and northeastern filament base cavity (Feature 3) are the most straightforward to analyze both spatially and spectrally—the former due to its large size and the latter due to the high number of available counts. Additionally, both cavities exhibit enhanced X-ray emission along their rims (see Figure \ref{fig:xray_features}). In the literature, X-ray bright rims are typically explained by two scenarios: supersonic expansion and buoyant uplift. In the first scenario, the bubbles expand supersonically, compressing and shocking the surrounding gas, forming a shocked shell of material around the bubbles \citep[e.g.,][]{fabian_deep_2003, forman_filaments_2007}. In the second scenario, the bubbles rise buoyantly in a subsonic phase of expansion, uplifting gas from the center as they rise  \citep[e.g.,][]{fabian_chandra_2000, churazov_evolution_2001, su_buoyant_2017, gendron-marsolais_uplift_2017}. 

\begin{figure}
\includegraphics[width=\linewidth]{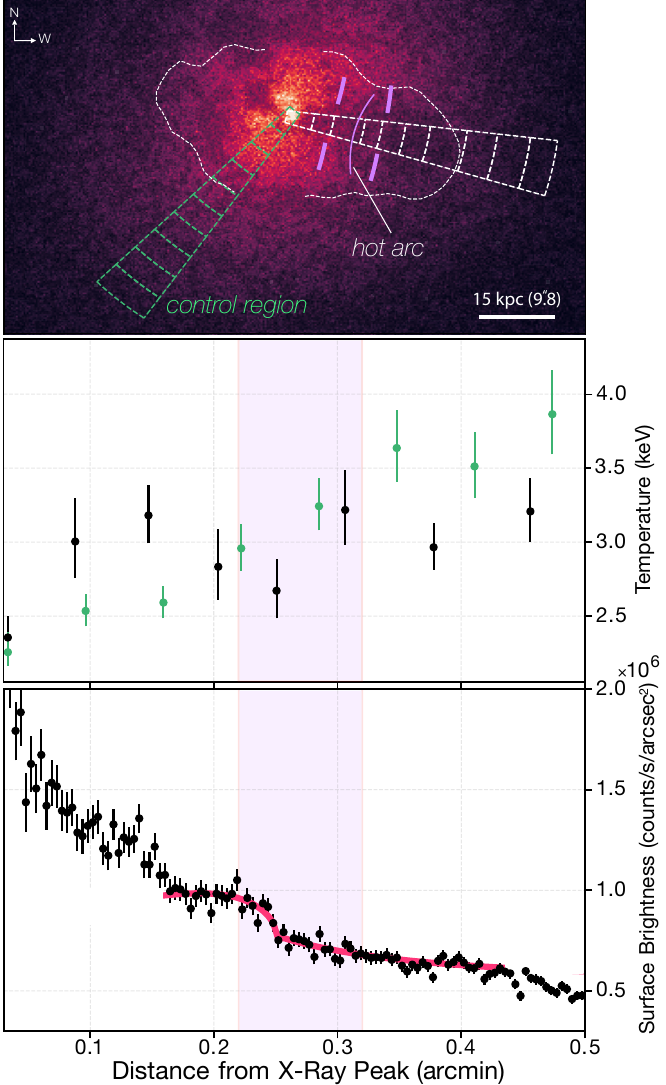}
\caption{Properties of the “hot arc.” \textit{Top:}  0\farcs5 binned \chandra\ image of the cluster core overlaid with a 9\farcs8$-$0\farcs98 unsharp mask to enhance surface brightness depressions. The hot arc identified in Figure \ref{fig:xray_tmap} is marked with a curved purple line. Spectral and surface brightness profiles are extracted across the southwestern cavity (dashed white boxes) and a control region rotated towards the southeastern sector of the core, chosen to avoid as many surface brightness fluctuations as possible. \textit{Middle:} Temperature profile across both regions, with green points corresponding to the control region and black points to the cavity region. The purple-shaded band denotes the region containing the hot arc. \textit{Bottom:} Surface brightness profile extracted across the cavity. A broken power-law model (red line) highlights a density discontinuity coincident with the hot arc.}
\label{fig:hotarc}
\end{figure}

\begin{figure*}
\centering
\includegraphics[width=\linewidth]{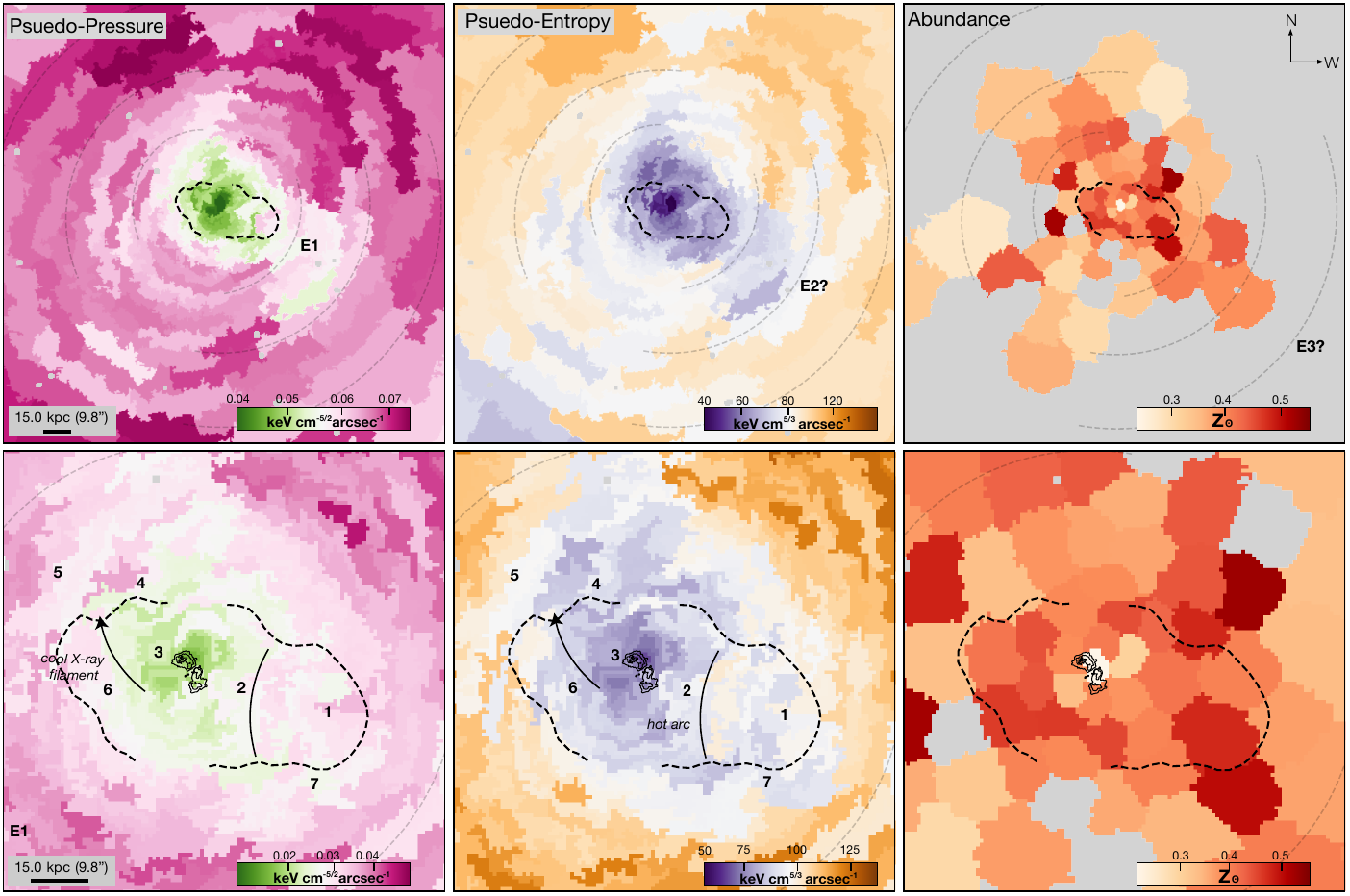}
\caption{Pseudo-pressure (\textit{left column}), pseudo-entropy (\textit{middle column}), and abundance (\textit{right column}) maps of the ICM in Abell 2597. \textit{Top row:} Central $\sim 310$ kpc view of each map. Lower-pressure and lower-entropy gas coincide with the coolest regions shown in Figure \ref{fig:xray_tmap}. The most metal-enriched gas is concentrated near the cluster core. \textit{Bottom row}: Central 60 kpc view of each map. Higher-pressure and higher-entropy gas surrounds the region just outside the 330 MHz radio source as well as the hot arc. The most enriched gas is primarily aligned with the axis of the 330 MHz source. }
\label{fig:xray_pezmap}
\end{figure*}

The northeastern filament base cavity contains slightly warmer gas and is fully surrounded by a cooler shell. The southwestern cavity shows a more complex structure, with the innermost cavity (Feature 2) surrounded by cooler gas. The outermost cavity (Feature 1) is generally surrounded by warmer gas, with hotter material, which \citetalias{tremblay_multiphase_2012} identified as a “hot arc,” separating it from the innermost cavity spectrally. Among the interpretations proposed in that work, two particularly relevant possibilities are: (1) the expanding cavity displaced cooler gas from the cluster center, creating the illusion of a hot arc when compared to adjacent cooler regions; or (2) the gas in the arc was locally heated as it rushed inward to refill the wake left by the buoyantly rising cavity, thermalizing cavity enthalpy.

To test these possibilities, we extracted a temperature profile across the hot arc and compared it to a control region rotated towards the southeast of the cluster core in order to avoid as many surface brightness fluctuations as possible (see top panel of Figure \ref{fig:hotarc}). Although the temperature varies across the southwestern cavity (middle panel), the hot arc region shows no significant excess compared to the control, supporting the interpretation that the apparent enhancement may result from displaced cooler gas rather than true heating. Additionally, our deeper data show no evidence of excess hard X-ray emission at the arc’s location, consistent with the idea that the feature may be a local soft X-ray deficit rather than a genuinely hot structure.

\begin{figure*}
\centering
\includegraphics[width=\linewidth]{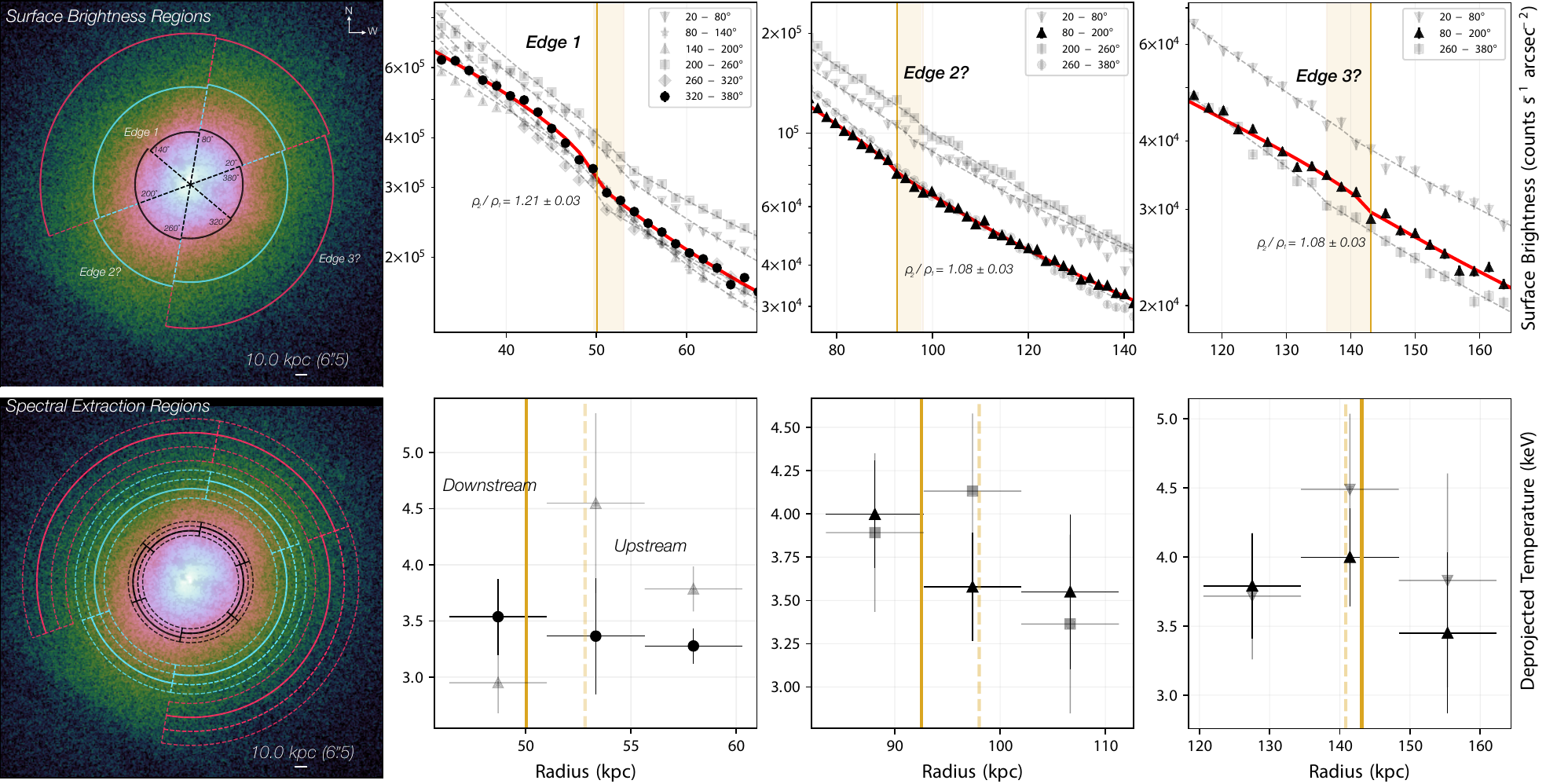} 
\caption{Surface brightness and deprojected temperature profiles across the candidate surface brightness edges in Abell 2597. \textit{Top left:} \chandra\ X-ray image showing the regions used to extract surface brightness profiles across each edge (black for Edge 1, light blue for Edge 2, and fuchsia for Edge 3), with solid lines marking sectors where an edge was identified from the broken power-law fits. \textit{Top right:} Surface brightness profiles extracted from each sector, with best-fitting broken power-law models overplotted as dashed grey lines and the example fit highlighted in solid red. Vertical yellow lines mark the edge position from the highlighted sector, while shaded yellow bands indicate the range of edge locations fit across all sectors. \textit{Bottom left:} \chandra\ image showing the regions used for spectral extraction, with solid lines corresponding to the approximate positions of each edge as determined from the surface brightness fits. \textit{Bottom right:} Deprojected temperature profiles derived from the spectral regions to the left, shown for two representative sectors per edge. In most cases, the temperature rises within $1-2\sigma$ just interior to the detected edge and declines immediately beyond it, consistent with the presence of weak shocks.}
\label{fig:cs_profiles}
\end{figure*}

However, our data suggest that the warm gas may not be confined to a narrow arc but could instead trace a broader, asymmetric rim encircling much of Cavity 1. If real, this morphology would resemble the shell-like structures seen around cavities undergoing supersonic expansion—an interpretation that \citetalias{tremblay_multiphase_2012} had previously disfavored due to the hot arc’s limited spatial extent. Since the strongest temperature enhancement occurs at the site of the hot arc,  we extracted a finely spaced surface brightness profile across the region (Figure \ref{fig:hotarc}, bottom panel) and fit it with a broken power law. The resulting density jump of $1.67 \pm 0.05$ corresponds to a Mach number of $\mathcal{M} \sim 1.47$, consistent with a weak shock.  If this shock is genuine, it could support a scenario in which the outer bubble expanded supersonically, compressing the surrounding ICM into a shell of heated gas. However, we note that the surface brightness enhancement has a concave-inward geometry, which is atypical for forward-moving shocks. Furthermore, our analysis assumes spherical symmetry in the shock model, which may not be strictly valid in this asymmetric region. Therefore, the derived Mach number should be interpreted cautiously.  

Figure \ref{fig:xray_pezmap} presents the pseudo-pressure, pseudo-entropy, and abundance maps of the ICM. The top row displays the large-scale structure, with pseudo-pressure and entropy distributions closely following the temperature map, and lower-pressure and lower-entropy gas coinciding with the coolest regions of the ICM. The abundance map, while more uncertain due to larger statistical errors ($\sim 20\%$), shows preferential enrichment along the radio-jet axis, and additional enrichment to the north. Across larger radii, we observe slightly elevated pressure and entropy at the locations of the surface brightness edges, consistent with weak shocks. However, we caution that the expected entropy increase from a $\mathcal{M} \sim 1.47$ shock is only $\sim$4\%—and even less for weaker shocks—which falls below the $\sim$10\% fractional uncertainties in our maps. As such, any real entropy jump would be difficult to confirm at this significance level. A similar trend is seen in the inner and middle edges of the abundance map, along the direction of the radio source.

Within the central 60 kpc (lower panels), the northeastern filament base cavity contains gas at higher pressure and entropy than its surroundings. On larger scales, the region enclosing the 330 MHz radio source is characterized by low pressure and entropy, with the hot arc exhibiting relatively elevated values in both. The abundance of gas within the extent of the radio source is also enhanced relative to the cluster mean ($\sim$0.4 $Z_\odot$).

\subsection{150 kpc - Scale Shocks?} 
\label{subsec:cocoon}

Cocoon shocks form as the expanding plasma from the AGN compresses and displaces the ambient gas \citep{enslin_reviving_2001}, manifesting as a surface brightness edge encircling the lobes.  Simulations have shown that as the shock ages (i.e., after only a few Myr), the typical radial velocity of cocoon material drops below the original launch speed, eventually below the sound speed of the ICM $c_s$, and the shock wave becomes a sound wave \citep{bourne_agn_2017, guo_reversing_2018}. If the surface brightness edges result from several successive shock waves that have gradually slowed down since launch, we should observe surface brightness discontinuities at the locations of each edge. 
\begin{table*}  
\hspace*{-3cm}
\resizebox{1.17\textwidth}{!}{
\begin{tabular}{lcccccc|c|ccccc}  
\hline  
\hline  
Edge & Sector & $r_J$ & $J$ & $\alpha_1$ | $\alpha_2$ & $\chi^2 | \ \mathrm{ d.o.f}$ & $\Delta \chi_\nu^2$ & C &$\mathcal{M}$ & $kT^{\rm d}$| $kT^{\rm u}$ & $p^d_{\rm ICM}$| $p^u_{\rm ICM}$ & $n^d_{\rm e}$| $n^u_{\rm e}$ & $E_{\rm shock}^{\rm total}$ \\
 & $(^\circ)- (^\circ)$ & (kpc) &  &  &  &  &  &  & (keV)  & ($10^{-10}$ erg cm$^{-3}$) & ($10^{-2}$ cm$^{-3}$)  & ($10^{59}$ erg) \\
\hline
1  & 20-80 & 53 & 1.11 $\pm$ 0.03 & 1.13 $\pm$ 0.05  & 28.52 & -4.55 & E & 1.07 $\pm$ 0.01 & 5.12 $\pm$ 0.85 & 6.36 $\pm$ 1.11 & 44.71 $\pm$ 7.60 & 7.2 \\
 & & & & 1.50 $\pm$ 0.05 & 17 & & & & 3.58 $\pm$ 0.16 & 7.75 $\pm$ 0.37 & 21.62 $\pm$ 1.00 & \\
 & 80-140 & 49 & 1.01 $\pm$ 0.03 & 1.32 $\pm$ 0.07  & 13.45 & -5.38 & E & 1.01 $\pm$ 0.01 & 3.54 $\pm$ 0.40 & 6.43 $\pm$ 0.78 & 23.94 $\pm$ 2.81 & 1.0 \\
 & & & & 2.28 $\pm$ 0.26 & 16 & & & & 2.70 $\pm$ 0.44 & 3.55 $\pm$ 0.64 & 22.73 $\pm$ 3.86 & \\
 & 140-200 & 52 & 1.21 $\pm$ 0.04 & 0.74 $\pm$ 0.11  & 31.70 & -7.65 & E & 1.14 $\pm$ 0.01 & 4.55 $\pm$ 0.80 & 5.30 $\pm$ 1.00 & 41.47 $\pm$ 7.53 & 11.5 \\
 & & & & 1.80 $\pm$ 0.36 & 18 & & & & 3.79 $\pm$ 0.20 & 7.12 $\pm$ 0.41 & 25.06 $\pm$ 1.37 & \\
 & 200-260 & 52 & 1.15 $\pm$ 0.03 & 1.24 $\pm$ 0.16  & 19.03 & -6.60 & E & 1.10 $\pm$ 0.01 & 4.36 $\pm$ 0.64 & 5.32 $\pm$ 0.86 & 38.48 $\pm$ 5.89 & 8.0 \\
 & & & & 1.38 $\pm$ 0.05 & 15 & & & & 3.55 $\pm$ 0.14 & 7.97 $\pm$ 0.34 & 20.86 $\pm$ 0.86 & \\
 & 260-320 & 51 & 1.16 $\pm$ 0.02 & 1.33 $\pm$ 0.06  & 35.78 & -2.57 & E & 1.10 $\pm$ 0.01 & 4.20 $\pm$ 0.45 & 6.78 $\pm$ 0.79 & 30.79 $\pm$ 3.43 & 10.9 \\
 & & & & 1.37 $\pm$ 0.03 & 20 & & & & 2.91 $\pm$ 0.60 & 3.65 $\pm$ 0.80 & 25.18 $\pm$ 5.32 & \\
 & 320-380 & 50 & 1.21 $\pm$ 0.03 & 0.60 $\pm$ 0.08  & 20.65 & -9.83 & E & 1.14 $\pm$ 0.01 & 3.54 $\pm$ 0.34 & 6.36 $\pm$ 0.67 & 24.14 $\pm$ 2.42 & 13.7 \\
 & & & & 1.73 $\pm$ 0.31 & 18 & & & & 3.36 $\pm$ 0.52 & 4.48 $\pm$ 0.76 & 28.07 $\pm$ 4.53 & \\
 \hline
 2  & 20-80 & 96 & 1.05 $\pm$ 0.04 & 1.18 $\pm$ 0.32  & 18.64 & 0.03 & NE & 1.03 $\pm$ 0.01 & 3.94 $\pm$ 0.49 & 2.43 $\pm$ 0.32 & 1.92 $\pm$ 0.09 & 8.4 \\
 & & & & 1.50 $\pm$ 0.05 & 15 & & & & 3.74 $\pm$ 0.58 & 1.93 $\pm$ 0.32 & 1.61 $\pm$ 0.09 & \\
 & 80-200 & 92 & 1.08 $\pm$ 0.03 & 1.10 $\pm$ 0.28  & 27.72 & -0.26 & U & 1.05 $\pm$ 0.01 & 4.00 $\pm$ 0.31 & 2.58 $\pm$ 0.22 & 2.02 $\pm$ 0.07 & 13.3 \\
 & & & & 1.54 $\pm$ 0.01 & 30 & & & & 3.58 $\pm$ 0.31 & 2.27 $\pm$ 0.21 & 1.98 $\pm$ 0.06 & \\
 & 200-260 & 98 & 1.04 $\pm$ 0.03 & 1.11 $\pm$ 0.24  & 41.78 & -0.13 & U & 1.03 $\pm$ 0.01 & 4.13 $\pm$ 0.45 & 2.94 $\pm$ 0.33 & 2.22 $\pm$ 0.08 & 8.3 \\
 & & & & 1.65 $\pm$ 0.02 & 28 & & & & 3.36 $\pm$ 0.52 & 1.64 $\pm$ 0.27 & 1.52 $\pm$ 0.10 & \\
 & 260-380 & 92 & 1.08 $\pm$ 0.02 & 1.42 $\pm$ 0.06  & 25.20 & -0.69 & U & 1.05 $\pm$ 0.01 & 3.54 $\pm$ 0.30 & 2.48 $\pm$ 0.22 & 2.18 $\pm$ 0.07 & 12.9 \\
 & & & & 1.63 $\pm$ 0.02 & 29 & & & & 3.55 $\pm$ 0.30 & 2.25 $\pm$ 0.20 & 1.98 $\pm$ 0.06 & \\
 \hline
 3  & 20-80 & 140 & 1.03 $\pm$ 0.03 & 1.68 $\pm$ 0.15  & 23.56 & 0.01 & NE & 1.02 $\pm$ 0.01 & 4.49 $\pm$ 0.55 & 1.52 $\pm$ 0.20 & 1.06 $\pm$ 0.04 & 8.9 \\
 & & & & 1.63 $\pm$ 0.04 & 26 & & & & 3.83 $\pm$ 0.78 & 1.02 $\pm$ 0.21 & 0.83 $\pm$ 0.03 & \\
 & 80-200 & 143 & 1.08 $\pm$ 0.03 & 0.95 $\pm$ 0.29  & 23.19 & -0.23 & U & 1.05 $\pm$ 0.01 & 4.00 $\pm$ 0.36 & 1.30 $\pm$ 0.12 & 1.01 $\pm$ 0.03 & 22.3 \\
 & & & & 1.66 $\pm$ 0.06 & 17 & & & & 3.45 $\pm$ 0.58 & 0.77 $\pm$ 0.14 & 0.70 $\pm$ 0.04 & \\
 & 260-380 & 136 & 1.06 $\pm$ 0.03 & 1.49 $\pm$ 0.09  & 22.76 & -0.36 & U & 1.04 $\pm$ 0.01 & 3.83 $\pm$ 0.34 & 1.43 $\pm$ 0.13 & 1.17 $\pm$ 0.04 & 19.4 \\
 & & & & 1.80 $\pm$ 0.11 & 22 & & & & 3.45 $\pm$ 0.32 & 1.20 $\pm$ 0.12 & 1.09 $\pm$ 0.03 & \\
\hline  
\end{tabular}  
}
\caption{Best-fitting parameters of the broken power-law models for the candidate surface brightness edges in Abell 2597, corresponding to the sectors shown in Figure~\ref{fig:cs_profiles}. Columns list the azimuthal range of each sector, the projected radius of the density jump ($r_J$), the jump amplitude ($J$), the inner and outer density slopes ($\alpha_1$, $\alpha_2$), and the $\chi^2$ value with degrees of freedom (d.o.f) for the broken power-law fit. The $\Delta\chi^2_\nu = \chi^2_{\nu,{\rm Bkn}}-\chi^2_{\nu,{\rm Single}}$ column gives the difference in reduced chi-squared between the broken and single power-law models (negative values favor the broken model). $C$ denotes the qualitative classification of each sector as a clear edge (E), uncertain (U), or not an edge (NE), based on whether the profile could also be fit by a single power law. The Mach number ($\mathcal{M}$) is derived from the best-fitting density jump using the Rankine–Hugoniot relation. The downstream ($d$) and upstream ($u$) quantities include the gas temperature ($kT$), pressure ($p_{\rm ICM}$), and electron density ($n_{\rm e}$), from which we estimate the corresponding shock energy ($E_{\rm shock}$).}
\label{tab:sb_edges}  
\vspace{-2mm}
\end{table*}  

We examined surface brightness profiles across each edge using wedges with opening angles ranging between $60-120^\circ$, shown in the top left panel of Figure \ref{fig:cs_profiles}. Within the central 150 kpc of the cluster, the \chandra\ PSF ranges between $\sim 0\farcs5 - 1\farcs0$. To minimize the influence of systematic artifacts, we extracted profiles using bin widths roughly twice as large as the PSF, corresponding to $\sim 1\farcs0$ for Edge 1, $1\farcs2$ for Edge 2, and $1\farcs5$ for Edge 3. Each radial bin contains $\sim700$–$5000$ counts. For Edge 3 in particular, the southeastern portion of the sector lies across a chip gap, which significantly reduces the count rate; we therefore excluded the $200$–$260^\circ$ azimuthal range from the analysis. 

The resulting surface brightness profile from each sector for each edge is shown in the top row of Figure \ref{fig:cs_profiles}. Each profile was fit with a broken power-law model using \textsc{PyProffit} \citep{eckert_low-scatter_2020}, which includes five parameters: a normalization constant, a density jump $J$, the radial position of the jump $r_J$, and the inner and outer slopes of the profile ($\alpha_1$ and $\alpha_2$). The results of these fits are summarized in Table \ref{tab:sb_edges}. When a significant surface brightness discontinuity is detected—indicating the presence of an edge—the best-fitting model is overplotted as a dashed line, with the sector yielding the most prominent discontinuity highlighted in red. To assess whether these discontinuities are statistically significant, we also fit single power-law models (results reported in Appendix \ref{app:spw_edges}) to each sector for comparison, with Table \ref{tab:sb_edges} listing the difference in reduced chi-squared, $\Delta\chi^2_\nu = \chi^2_{\nu,{\rm Bkn}}-\chi^2_{\nu,{\rm Single}}$, between the two models.

For Edge 1, the reduced $\chi^2$ values for the single power-law fits consistently exceed $\sim4$ in every sector, while the broken power-law fits yield reduced $\chi^2$ values close to unity. This substantial improvement indicates that the broken power-law model provides a significantly better description of the data, strongly supporting the presence of a true surface brightness edge. In contrast, for Edges 2 and 3, the broken power-law fits only marginally outperform the single power-law models, with both yielding reduced $\chi^2 \approx 1$. As a result, these features cannot be confidently classified as edges. Most sectors are therefore labeled as “uncertain” (U), while the $20$–$80^\circ$ sectors for both Edges 2 and 3 favor the single power-law model and are classified as “not an edge” (NE). The weaker appearance of these outer edges likely reflects their intrinsically lower surface brightness contrast, further smoothed by the broader annular widths used at larger radii ($1\farcs25$–$1\farcs5$) to account for the varying \chandra\ PSF. 

Although Edge 1 appears the most circular, its density jump varies modestly with azimuth, ranging from $1.01 \pm 0.01$ in the $80$–$140^\circ$ sector, which overlaps with the northeastern ghost cavity, to $1.21 \pm 0.03$ in the $320$–$380^\circ$ sector. Comparable azimuthal variations are observed for Edges 2 and 3, with density jumps ranging from $1.04 \pm 0.03$ $-$ $1.08 \pm 0.03$ for Edge 2 and $1.03 \pm 0.03$ $-$ $1.08 \pm 0.03$ for Edge 3. 

We examined the thermodynamic properties across each edge by extracting spectra from the wedges shown in the lower left panel of Figure \ref{fig:cs_profiles}, each containing approximately $9,000$–$15,000$ counts. The right panels show the resulting deprojected temperature profiles, derived with the \textsc{Deproject} package in \textsc{CIAO}/\textsc{Sherpa}, for two representative sectors per edge. The downstream regions correspond to the compressed, heated gas just inside the edge, while the upstream regions trace the undisturbed ICM immediately outside. The downstream and upstream temperatures for each sector are listed in Table \ref{tab:sb_edges}. Using the deprojected temperatures, we also calculated the deprojected pressures and electron densities in each wedge. 

Despite the presence of numerous X-ray cavities within the central region introducing substantial morphological complexity, we observe consistent temperature rises across Edge 1; downstream values exceed upstream values, but this rise is generally consistent within errors and occasionally reaches $1$–$2\sigma$ significance. Only the $80$–$140^\circ$, $260$–$320^\circ$, and $320$–$380^\circ$ sectors exhibit corresponding pressure increases across the edge, as would be expected for a weak shock front. For Edges 2 and 3, the temperature uncertainties are larger, and although downstream temperatures are generally higher than upstream values (except in Edge 2’s $260$–$380^\circ$ sector), both sides are typically consistent within $1\sigma$. Similarly, while the pressures are systematically higher on the downstream side than on the upstream side, their uncertainties largely overlap. This raises the possibility that if Edges 2 and 3 are genuine discontinuities, they may represent cold fronts rather than shocks. However, interpreting them as cold fronts also presents challenges, which we discuss further in Section \ref{subsec:agn_history}.

For now, we interpret the edges as shock fronts and estimate their Mach numbers, $\mathcal{M}$, based on the fitted density jumps, since the density contrasts are detected at much higher significance than the downstream temperature and upstream temperatures. Using the Rankine-Hugoniot relation, we can correlate each density jump $J$ to its Mach number with:
\begin{equation}
\mathcal{M} = \left(\frac{3J}{4-J}\right)^{1/2}.
\end{equation}
We obtain Mach numbers ranging from $1.07 \pm 0.01 - 1.14 \pm 0.01$, $1.03\pm 0.01 -1.05 \pm 0.01$, and $1.02\pm 0.01 -1.05 \pm 0.01$ for the inner, middle, and outer edges, classifying each as very weak shock fronts. 

Following Section 5.6 of \citet{randall_shocks_2011}, we can estimate the total energy $E$ imparted to the ICM from each shock as:
\begin{equation}
E \approx PV \left( f_P - 1 \right),
\end{equation}
\noindent where $V$ is the gas volume contained within the shock front, $P$ is the pressure across the downstream region of the front, and $f_P = (P + \Delta P )/P$ describes the ratio of the post- and pre-shock pressure. We can relate $f_P$ to the Mach number from the Rankine-Hugoniot shock jump conditions as 
\begin{equation}
f_P  = \frac{2 \gamma \mathcal{M}^2 - (\gamma - 1)}{\gamma + 1},
\end{equation}
\noindent where $\gamma = 5/3$.

\begin{table*}
\centering
  \begin{tabular}{cccccccccc}
    \hline\hline
    Label & Cavity & $R$ ($D$) & $r_{major}$ & $r_{minor}$ & $v_{\rm term}$ & $t_{\rm buoy}$  & $t_{\rm c_s}$ & $pV$ & $P_{\rm cav}$\\
     &  & (kpc) & (kpc) & (kpc) & (\kms) & ($10^7$yr) & ($10^7$yr) & ($\times 10^{57}$ erg) & ($\times 10^{43}$ erg s$^{-1}$) \\
    (1) & (2) & (3) & (4) & (5) & (6) & (7) & (8) & (9) & (10)\\
    \hline
1 & Inner SW Ghost Cavity & 9.6 & 7.6 & 7.6 & 827 & 1.1 & 1.2 & 41.0 & 44.9 \\
2 & SW Ghost Cavity & 26.2 & 7.6 & 7.6 & 502 & 5.1 & 3.2 & 21.1 & 6.4 \\
3 & NE Filament Base Cavity & 8.0 & 3.3 & 3.3 & 599 & 1.3 & 1.0 & 3.7 & 4.1 \\
4 & NE Ghost Cavity & 21.1 & 7.0 & 7.0 & 539 & 3.8 & 2.6 & 19.8 & 7.8 \\
5 & Outer NE Ghost Cavity & 37.1 & 8.0 & 8.0 & 433 & 8.4 & 4.5 & 18.8 & 3.7 \\
6 & Eastern Ghost Cavity & 16.1 & 6.7 & 3.5 & 434 & 3.6 & 2.0 & 5.6 & 2.5 \\
7 & Broken off Pieces  & 29.4 & 15.4 & 4.9 & 381 & 7.5 & 3.6 & 16.4 & 3.7 \\
 & of SW Ghost Cavity? &  &  &  &  &  & &  &  \\
\hline
  \end{tabular}
   \caption{Properties of the X-ray Cavities in Abell 2597. Columns: (1) - (2) Cavity label and name given in Table \ref{tab:xray_features}; (3) Projected distance from the X-ray peak to the center of the cavity; (4) - (5) Major and minor radii; (6) Terminal velocity of the buoyantly rising bubble (7) - (8) Age assuming rise at terminal velocity / sound speed; (9) $pV$ work done by inflation; (10) Cavity power, assuming relativistic plasma filling.}
  \label{tab:xray_cavity_measurements}
\end{table*}

To estimate the volume of shocked gas, we assume that the entire region enclosed by each shock front has been affected by the passage of the shock. We model the volume enclosed by Edge 1 as a sphere with radius $50 \rm \ kpc$, and the volumes enclosed by Edges 2 and 3 as prolate ellipsoids, with the semi-major and semi-minor axes of 92 and 98 kpc for Edge 2, and 136 and 143 kpc for Edge 3. A range of values for the total shock energy for each edge is then computed using the Mach numbers and downstream pressure reported in each sector. We find energies ranging from $\sim 1 - 22 \times 10^{59}$ erg per front.  Because we assume a constant Mach number over each shock's life, the energy calculated using this formula may slightly underestimate the total shock energy, as the Mach number likely would have been higher during earlier stages of the shock’s development.  Nevertheless, the shock energy derived from this approach generally aligns within a factor of a few with estimates from hydrodynamical simulations of a central point explosion (e.g., \citealt{randall_shocks_2011}).

\subsection{X-Ray Cavities}
\label{subsec:cavities}

To estimate the energetics of the cavities identified in Section \ref{subsec:xray_features}, we estimate their buoyant rise times and the energy they dissipate as they ascend. The results, along with each cavity’s effective radius—spherical or elliptical as appropriate—are summarized in Table \ref{tab:xray_cavity_measurements}.

Assuming each cavity formed near the X-ray peak, where the active radio source is located, and rose buoyantly to their current projected positions, we can estimate the cavity age as 
\begin{equation}
t_{\text{cav}} = R/v,
\end{equation}
\noindent where $R$ is the projected distance from the ICM peak and $v$ is the bubble’s rise velocity. To estimate $v$, we consider two characteristic velocities: the sound speed $c_s$, which serves as an upper limit on gas motions within the ICM, and the terminal velocity $v_t$, which accounts for the impact of the ICM's ram pressure on the buoyancy force. 

We estimate $c_s$ using 
\begin{equation}
c_s = \sqrt{\gamma kT / \mu m_p},
\end{equation}
\noindent where the adiabatic index for a monatomic gas $\gamma = 5/3$, $m_p$ is the proton mass, and the mean molecular weight $\mu = 0.62$. The terminal velocity is calculated following \cite{churazov_evolution_2001}:
\begin{equation}
v_t = \sqrt{\frac{2gV}{SC}},
\end{equation}
\noindent where $V$ is the bubble's volume, $S$ is its cross-sectional area, $C = 0.75$ is the drag coefficient, and $g$ is the gravitational acceleration. 

Each cavity is modeled as a prolate sphere or ellipsoid, with the minor axis radius taken as the depth (see Table \ref{tab:xray_cavity_measurements}). Following \citet{birzan_systematic_2004}, we calculate $g$ using the stellar velocity dispersion of the BCG, under the approximation that the galaxy is an isothermal sphere: $g \approx 2\sigma^2/R$, where $\sigma = 350$ km s$^{-1}$ is the median stellar velocity dispersion across the BCG (see Figure 11 in \cite{tremblay_galaxy-scale_2018}) and $R$ is the projected distance of the cavity from the center of the BCG. 

As cavities rise, they do work on the surrounding medium. We can estimate the cavity power, $P_{\rm cav}$, as the ratio of the cavity enthalpy $H_{\rm cav}$ to its age $t_{\rm cav}$ as 
\begin{equation}
P_{\rm cav} = H_{\rm cav}/t_{\rm cav} = 4pV/t_{\rm cav},
\end{equation}
\noindent where $p$ is the ambient ICM pressure, $V$ is the cavity volume, and $t_{\rm cav}$ is the average of the cavity ages estimated using both the terminal velocity ($t_{\rm buoy}$) and the sound speed ($t_{\rm c_s}$) reported in Table \ref{tab:xray_cavity_measurements}.

Using the surrounding pressures reported in Table \ref{tab:xray_cavity_measurements}, we obtain cavity enthalpies ($4pV$) ranging from $\sim 2-20 \times 10^{58}$ erg.  Using the same average age above, we calculate cavity powers in the range of $4-20 \times 10^{43}$ erg s$^{-1}$. We discuss how this feedback counteracts the ongoing cooling flow in Abell 2597 further in Section \ref{subsubsec:cav_heat}.

\subsection{The X-Ray Channel}
\label{subsubsec:xray_channel}

\begin{figure}
\hspace{-2mm}
\includegraphics[width=\linewidth]{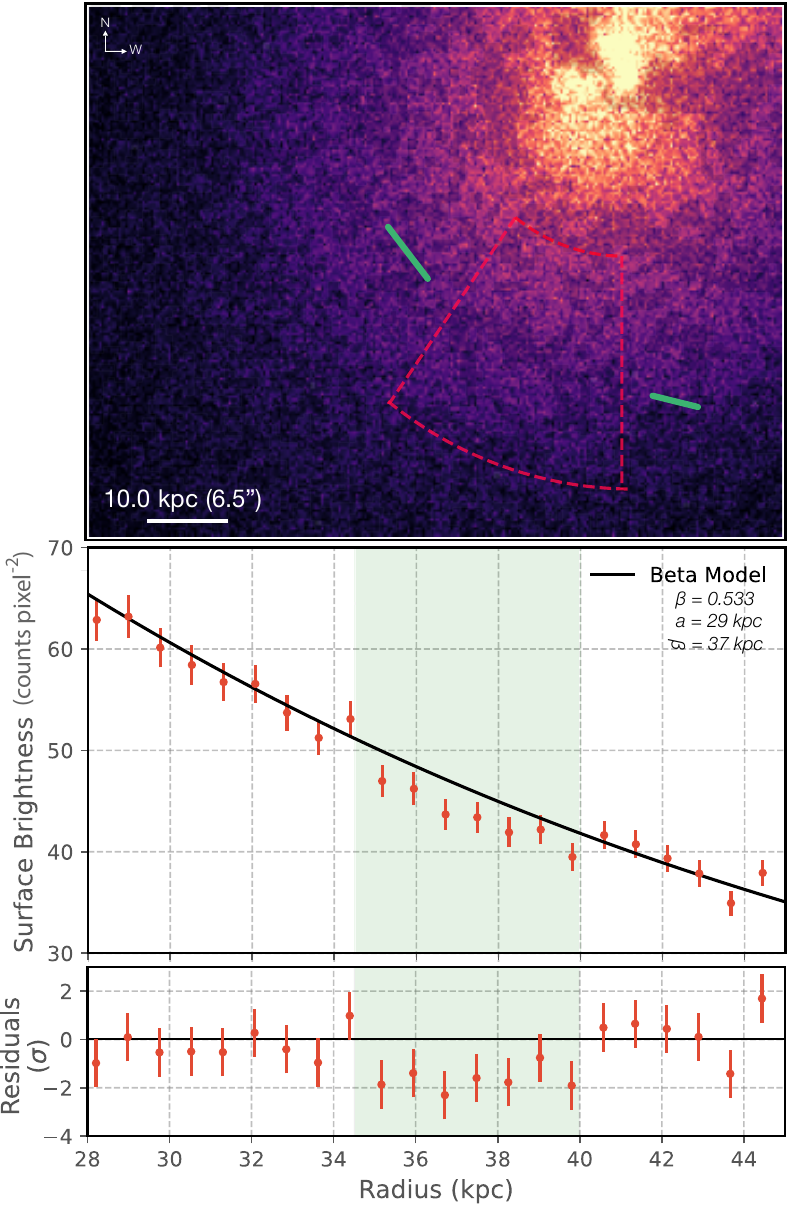}
\vspace{-2mm}
\caption{\textit{Top:} Unfiltered X-ray image overlaid with a 9\farcs8 - 0\farcs98 unsharp mask to enhance surface brightness depressions. The outer edges of the X-ray channel are marked with solid green lines. Finely spaced surface brightness profiles are extracted from the wedge outlined by dashed red lines. \textit{Bottom:} Surface brightness profile fitted with a 1D beta model assuming no deficit. The channel (green shaded region) is detected at $\sim 1-2\sigma$ below the model, indicating a $\sim 10\%$ brightness deficit.}
\vspace{-3mm}
\label{fig:sb_channel_profile}
\end{figure}

The X-ray ``channel,'' shown again in the top panel of Figure \ref{fig:sb_channel_profile}, runs roughly parallel to the southwestern radio jet and is located a distance of $\sim 35$ kpc to the southeast of the X-ray peak. The feature extends $\sim 57$ kpc ($37\farcs0$) in length and $\sim 8$ kpc ($5\farcs3$) in width.  The lower panel of Figure \ref{fig:sb_channel_profile} presents the surface brightness profile extracted across the channel. To quantify deviations from the expected gas distribution, we fit the radial profile with a 1D $\beta$-model in \textsc{CIAO}, excluding the data points corresponding to the channel region from the fit. A $\sim 1-2\sigma$ deviation is observed at the location of the channel, corresponding to a significant local deficit relative to the model.

\begin{figure*}
\hspace{-2mm}
\includegraphics[width=\linewidth]{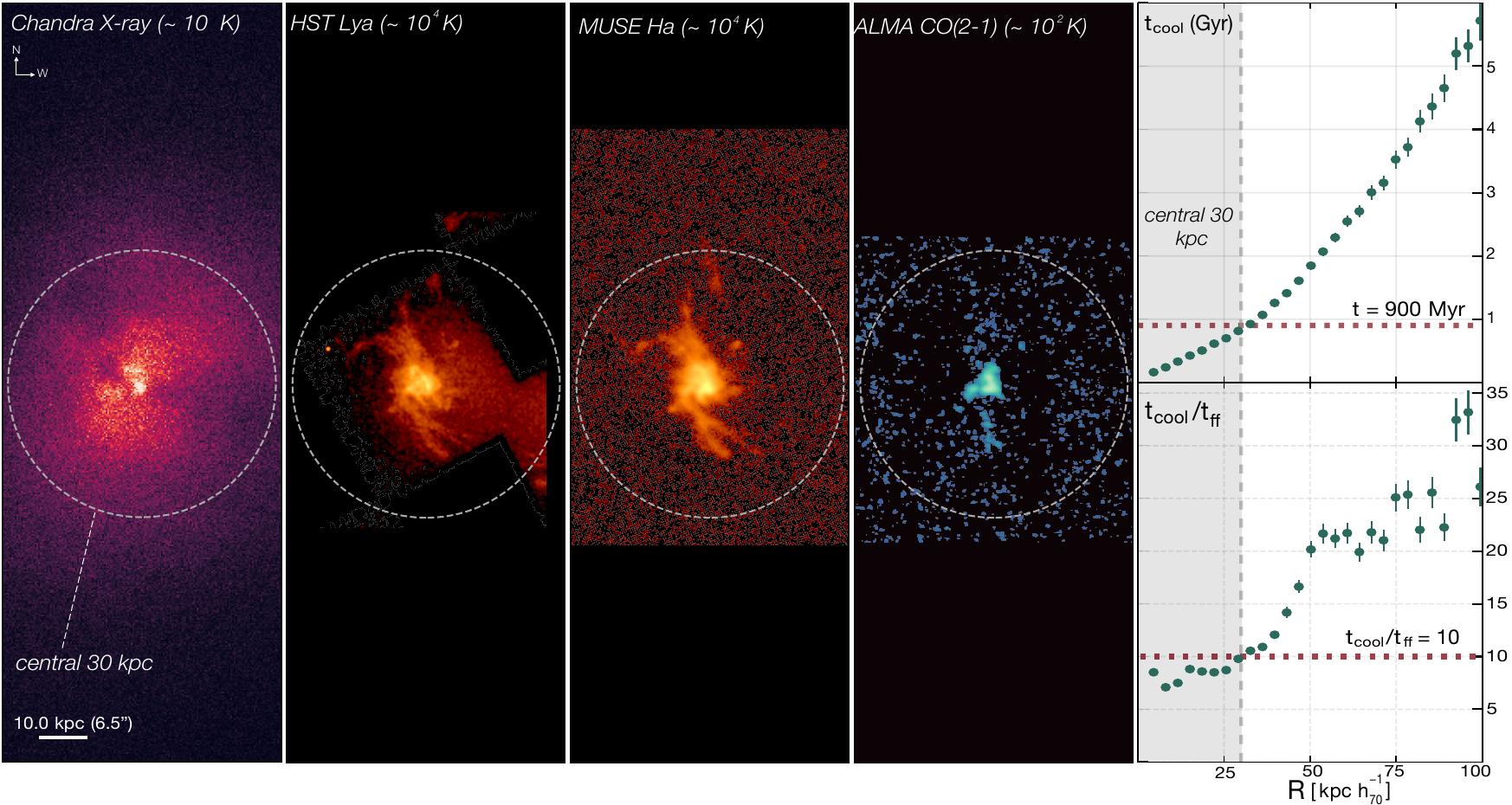}
\vspace{-2mm}
\caption{\textit{Left four panels:} Multi-wavelength gallery from X-ray to sub-mm illustrating the multiphase gas within the central 30 kpc (dashed circle) of Abell 2597. \textit{Right:} Radial timescales characterizing the thermodynamic state of the ICM, with the top panel showing the cooling time $t_{\text{cool}}$, and the lower panel displaying cooling-to-free-fall time ratio $t_{\text{cool}} / t_{\text{ff}}$. }
\label{fig:timescales}
\end{figure*}

Such channels have been interpreted as thin sheets of lower-density gas embedded within the ICM \citep{markevitch_shocks_2007}.  Assuming approximate spherical symmetry for the main cluster body and a complete absence of gas within the three-dimensional channel structure, we can estimate the minimum line-of-sight extent, $d$, required to produce the observed projected X-ray brightness deficit.

The X-ray surface brightness at a projected radius $\varpi$ is given by
\begin{equation}
S_X(\varpi) = \int_{-\infty}^{\infty} n_e^2(r) \, dz.
\end{equation}
For a $\beta$-model density profile, where $n_e(r) \propto (1 + r^2/a^2)^{-3\beta/2}$, this integral evaluates to
\begin{equation}
S_X(\varpi) = n_{e,0} n_{p,0} a (1 + \varpi^2/a^2)^{-3\beta+1/2} B(3\beta - 1/2, 1/2),
\end{equation}
where $B(x,y)$ is the Beta function.

A cavity or lower-density channel removes part of this emission measure.  If the channel extends over a depth $d$, the missing emission is:
\begin{equation}
\delta S_X(\varpi) = \int_{-d/2}^{d/2} n_e^2(r) \, dz.
\end{equation}
Approximating $n_e^2(r)$ as constant over this small depth, the missing emission is simply $\delta S_X(\varpi) \approx n_e^2(r) d$, giving the fractional deficit:
\begin{equation}
\frac{\delta S_X(\varpi)}{S_X(\varpi)} = \frac{d}{\sqrt{\varpi^2 + a^2} B(3\beta - 1/2, 1/2)}.
\end{equation}

Using the best-fit parameters for the model shown in the lower panel of Figure \ref{fig:sb_channel_profile}, with $\beta = 0.533$, $a = 29$ kpc), and an observed 10\% surface brightness deficit ($\delta S_X / S_X = 0.1$), we find that the minimum depth of the channel is $\sim 9$ kpc. Since the channel likely retains some gas, this depth represents only a lower limit, and the true extent could be significantly greater. Similar subtle X-ray surface brightness channels have been observed in the merging cluster Abell 520 \citep{wang_merging_2016} and in the sloshing cool core of Abell 2142 \citep{wang_deep_2018} with even greater estimated depths ($\sim 35$ kpc), suggesting that such features may be a common consequence of gas motions in dynamically evolving clusters. We discuss potential formation scenarios for this structure in Section \ref{subsec:origin_channel}.

\section{Heating the Rapidly Cooling ICM}
\label{sec:heating_icm}

\subsection{Thermal Instabilities and Residual Precipitation}
\label{subsec:thermal_instability}

In the classic picture of ICM cooling, thermal instabilities drive the condensation of colder, denser gas phases. As the ICM radiates energy, certain regions become unstable, leading to localized runaway cooling rather than uniform cooling across the cluster \citep{mcnamara_mechanical_2012}.

Several key timescales are predicted to govern the onset of thermal instabilities within the ICM. The free-fall time, $t_{\rm ff}$, describes the time required for a gas parcel to fall from rest to the center of the system under the gravitational potential of the cluster. In idealized thermal instability models, local density fluctuations can condense out of the hot phase when the ratio $t_{\rm cool}/t_{\rm ff}$ drops below a critical threshold, typically $\sim$10 in clusters \citep{gaspari_cause_2012, mccourt_thermal_2012, sharma_structure_2012}. The lower right panel of Figure \ref{fig:timescales} shows that the ratio falls below 10 within the central 30 kpc of Abell 2597, consistent with the presence of multiphase gas observed in the region.

To quantify the cooling rate, the upper right panel of Figure \ref{fig:timescales} plots the cooling time $t_{\rm cool}$ as a function of the radius within the cluster. We estimated $t_{\rm cool}$ using
\begin{align}
t_{\text{cool}} (r)\equiv \frac{3}{2} \frac{(n_e + n_p) kT(r)}{n_en_H \Lambda(kT(r), Z)},
\label{eqn:tcool}
\end{align}
\noindent where $\Lambda$ is the cooling function \citep{sutherland_cooling_1993}. We evaluated $\Lambda$ using an analytic fit \citep[][Table 3]{tozzi_evolution_2001}, calibrated with coefficients describing gas with average metallicity $Z=0.3 Z_\odot$. With  $r_{\rm cool}$  defined as the radius where  $t_{\rm cool} = 3$  Gyr \citep{mcdonald_revisiting_2018}, the classical cooling rate within the central $r_{\rm cool} \sim 60$ kpc is $\sim 210$ \mrate. 

Observations, however, suggest lower cooling rates.  Using \textit{XMM-Newton} Reflection Grating Spectrometer data, \citet{fabian_hidden_2023} estimated an X-ray cooling rate of $\sim 67^{+139}_{-19}$ \mrate\ after correcting for intrinsic absorption by cold gas and dust in the cluster core. This value is consistent with rates inferred from \ovi\ ultraviolet emission detected by FUSE, which—assuming the line traces gas cooling through $\sim 7 \times 10^5$ K—yields a $\sim 40$ \mrate\ cooling rate within $\sim$46 kpc, or up to $150^{+140}_{-100}$ \mrate\ after correcting for dust extinction \citep{oegerle_fuse_2001}.

Recent HST COS observations, which provide higher resolution UV data than FUSE, also detected the \ovi\ line, but within the central 2 kpc of the cluster (Vaddi \& Omoruyi et al., in prep). From the line’s luminosity, Vaddi \& Omoruyi et al., in prep estimate a cooling rate of $15.6 \pm 0.9$ \mrate\ within the nuclear region. Assuming this rate holds roughly constant out to $\sim 30$ kpc, where ALMA measures a cold gas mass of $3.2 \pm 0.1 \times 10^9$ \msun\ \citep{tremblay_cold_2016, tremblay_galaxy-scale_2018}, the cold gas reservoir could be replenished within the $\sim 190$ Myr cooling time at that radius. Since the UV-based cooling rate only corrects for galactic dust extinction, it likely represents a lower limit, further supporting ongoing residual cooling as a source of the cold gas.

\subsection{AGN Feedback History and Power}
\label{subsec:agn_history} 
\begin{figure*}
\includegraphics[width=\linewidth]{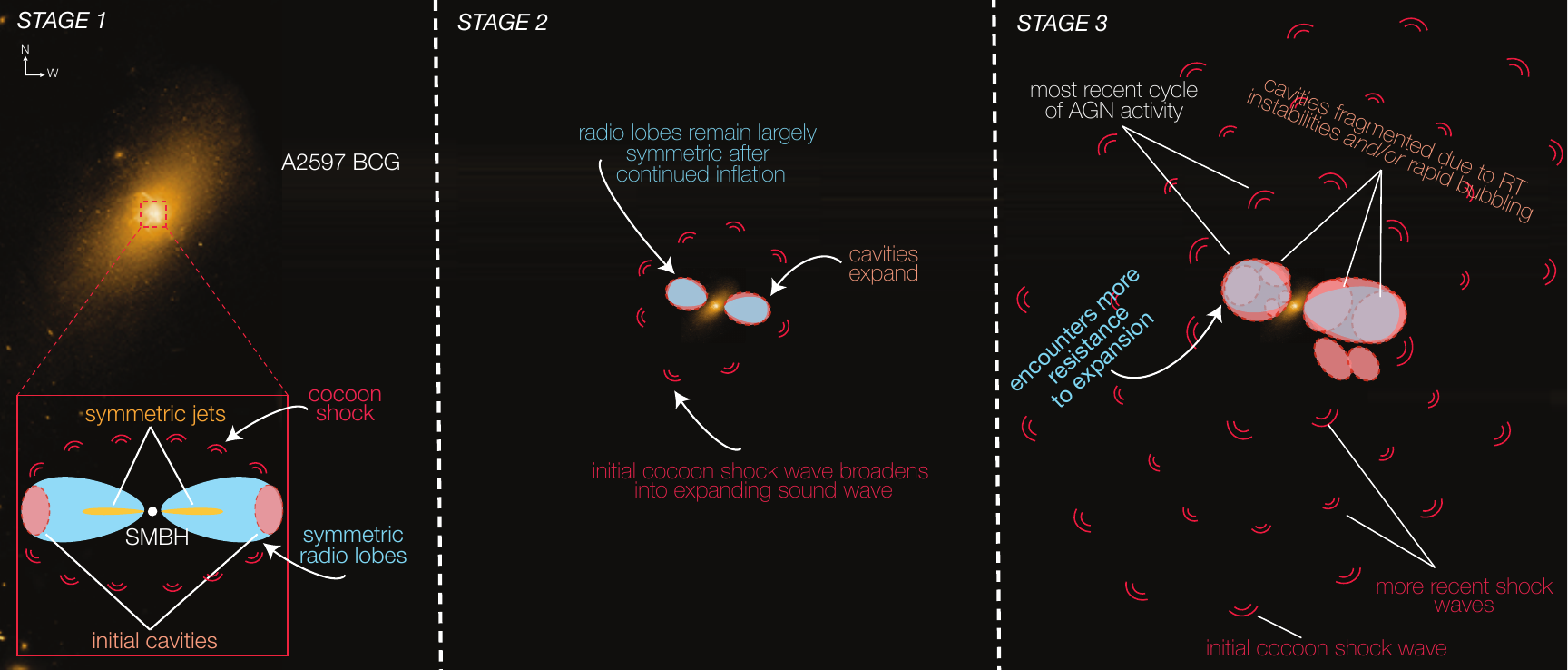}
\caption{Cartoon illustrating our interpretation of the features observed in Abell 2597’s ICM. Initially, symmetric jets inflate X-ray cavities and drive a cocoon shock into the ICM (Stage 1). As the lobes expand, the shock weakens and transitions into a sound wave (Stage 2). Repeated outbursts can produce multiple concentric shock fronts and cavities over time (Stage 3). The clearly detected surface brightness edge (Edge 1), and the possible additional edges (Edges 2 and 3), if real, likely trace successive shocks driven by multiple episodes of AGN activity. However, it remains unclear whether the observed cavities were also produced in separate episodes, or during the most recent outburst, with Rayleigh-Taylor instabilities fragmenting an initially large cavity pair into the observed structures.}
\label{fig:cartoon}
\end{figure*}

The most likely explanation for reducing the $\sim$210\mrate\ classical cooling flow to the observed $\sim$15\mrate\ residual rate is AGN feedback. As shown in Section \ref{sec:overall_icm}, Abell 2597 hosts a complex network of X-ray cavities and up to three surface brightness discontinuities likely associated with weak shocks. Together, these features suggest a multi-stage feedback cycle in which radio jets inflate cavities, drive shocks, and redistribute energy throughout the ICM \citep{mcnamara_mechanical_2012, fabian_observational_2012}. Whether these structures resulted from multiple discrete AGN outbursts separated by long quiescent periods, or a single prolonged episode of feedback, remains an open question. We explore both possibilities in the sections that follow and provide a schematic of our interpretation in Figure \ref{fig:cartoon}.

\subsubsection{Multiple Outbursts or Continuous Bubbling?}
\label{subsubsec:cav_origin}

The presence of multiple cavities located at a variety of distances from the central AGN suggests recurrent AGN activity, with each outburst leaving a distinct imprint on the surrounding gas. However, since the oldest cavity (Cavity 2) has a buoyant rise time of only $\sim$50 Myr, \citetalias{clarke_low-frequency_2005} suggested that none of the observed cavities may be true ``ghost” cavities, but rather products of the ongoing AGN feedback episode that has launched bubbles in several directions. Determining whether the radio plasma filling the cavities originates from distinct outbursts or a single sustained event requires tracing spectral age gradients. However, the limited frequency coverage and sensitivity of both past radio observations and the archival GMRT data used in this work make this infeasible.

An alternative approach to constrain the timing between each feedback episode is to estimate the ages of the surface brightness edges identified. However, the physical nature of these edges, particularly Edges 2 and 3, remains uncertain. Edge 1 appears to be a genuine weak shock, with thermodynamic properties consistent with gas compression and heating across the front. Although Edges 2 and 3, if real, show thermodynamic structure broadly consistent with weak shocks, their temperature and pressure jumps are typically constant within errors across each sector, so they could instead represent isothermal shocks \citep[e.g.][]{fabian_very_2006} or cold fronts. 

Cold fronts typically arise from contact discontinuities where cool, dense gas moves through a hotter medium, producing sharp, arc-like density jumps \citep[see review by][]{markevitch_shocks_2007}. In our case, although the outer two edges appear elliptical in projection, the measured density discontinuities are only prominent in three out of four sectors for Edge 2, and two out of three sectors for Edge 3. This localized measurement opens the possibility that the fronts are asymmetric or confined to specific sectors, which would be more consistent with a cold front origin. However, cold fronts are typically created by mergers, and Abell 2597 still appears dynamically relaxed given the spatial coincidence of the X-ray peak with the BCG optical center. While simulations show that residual sloshing motions from a past major merger can persist for several billion years \citep{zuhone_sloshing_2011}, it seems unlikely that they would produce two fronts both nearly elliptical in projection and with the observed thermodynamic properties. We therefore prefer the shock interpretation.

Assuming all three features are true edges and indeed shocks, the time required for the shocks to reach their current positions, $r_{\text{sh}}$, can be estimated using their Mach numbers from Section \ref{subsec:cocoon} with: 
\begin{equation}
t_{\text{age}} = \frac{r_{\text{sh}}}{\mathcal{M} c_S},
\end{equation}
\noindent where $c_s$ is the sound speed of the undisturbed gas ahead of the shock, determined using the upstream temperature. It is important to note that this approach may slightly overestimate the actual age of the shock by about 10--20\% \citep[as noted by][]{russell_chandra_2010}, since the shocks likely started with higher Mach numbers when first launched. Furthermore, the shocks are elliptical and slightly asymmetric, meaning their evolution may vary across different regions of the ICM. Furthermore, if projection effects are significant, the true 3D distance traveled by the shocks would exceed the projected radius, such that $r_{\text{sh}}$ is underestimated and the inferred $t_{\text{age}}$ is correspondingly biased low.

Applying this approach, we estimate the ages of the shock fronts to be approximately 46 Myr for the inner shock at 50 kpc (sector 320-380$^\circ$), 89 Myr for the middle shock at 92 kpc (sector 80-200$^\circ$), and 141 Myr for the outer shock at 143 kpc (sector 80-200$^\circ$). Their ages imply an average cycle time of $\sim 5 \times 10^7$ years in between outbursts. Remarkably, this repetition timescale is consistent with that observed across a wide range of systems, from Seyfert galaxies like NGC 2639 \citep{rao_agn_2023}, galaxy groups like NGC 5813 \citep{randall_shocks_2011} to clusters such as MS 0735.6+7421, Perseus, and Hydra A \citep{vantyghem_cycling_2014}, and also in higher-redshift systems like RBS 797 \citep{ubertosi_multiple_2023}. This suggests that $\sim 10^7$ year AGN outburst cycles may be a common feature of self-regulated feedback across a broad range of group and cluster environments. 

Given that the oldest cavity is approximately the same age as the innermost shock, and that the inferred AGN duty cycle is comparable to the age of this cavity, it's possible that all the observed cavities were produced during a single, ongoing episode of AGN feedback rather than by multiple discrete outbursts.  One can imagine that the observed bubbles are not all discrete structures but rather fragmented remnants of a single major outburst. 

In this scenario, the most recent cycle of AGN activity initially produced a bipolar cavity pair to the southwest and northeast, but Rayleigh-Taylor instabilities broke them into multiple smaller structures as they expanded and interacted with the surrounding ICM \citep{bruggen_buoyant_2001, pizzolato_rayleigh-taylor_2006, mcnamara_heating_2007}. Rayleigh-Taylor instabilities occur when a lighter fluid (e.g., the relativistic plasma in the cavity) accelerates into a denser one (e.g., the surrounding ICM), leading to the growth of perturbations along the interface. As bubbles rise buoyantly through the ICM, they are especially susceptible to these instabilities, which can fragment or shear the cavities over time. This process would naturally explain the asymmetries within Cavity 7, which does not resemble a typical spherical cavity and appears to have once been connected to Cavities 1 and 2. 

If a similar process occurred to the northeast, the observed depressions comprising Cavities 3, 4, 5 and 6 could actually be the fragmented remains of an initially larger cavity. The asymmetry of the jets, likely influenced by variations in ICM density and pressure, may have caused the northeastern cavities to fragment more rapidly than those to the southwest, where the southwestern lobe appears to have expanded more freely.

Similar cavity breakup and disruption have been observed in other systems, such as Abell 2052, where the rim of the northern cavity appears to be breaking apart as radio plasma leaks out (see Figure 4 in \cite{blanton_active_2010}; also \citet{blanton_chandra_2001, blanton_very_2011}). M87 also shows signs of instability-driven fragmentation, with a chain of three closely-spaced bubbles extending along the axis of the ``bud” cavity \citep{forman_reflections_2005}. The regular spacing and alignment of these bubbles suggest they likely formed during the same cycle of AGN activity, and imply that cavities can fragment without being completely disrupted, retaining coherent structure even as they evolve \citep{mcnamara_heating_2007}. 

Under this interpretation for A2597, the outer two shock edges, if real, were likely driven by earlier AGN outbursts, while the innermost edge corresponds to the current episode. 

\subsubsection{Inefficient Heating}
\label{subsubsec:cav_heat}

Regardless of the exact AGN feedback history, the injected energy should be more than sufficient to counteract radiative cooling. At an average cluster temperature of 3.5 keV and cooling radius of $r \sim 60$ kpc, the total cooling luminosity $L_{\rm cool} \approx 2 \times 10^{44}$ erg s$^{-1}$ matches the total power output of the expanding X-ray cavities, $P_{\rm cav} \sim 2 \times 10^{44}$ erg s$^{-1}$ (see Table \ref{tab:xray_cavity_measurements}). This balance, first noted in \citetalias{tremblay_multiphase_2012} and confirmed here, suggests AGN feedback regulates cooling in the cluster core, consistent with trends in nearly all other nearby cool-core clusters, with Abell 2597 fitting neatly along the $P_{\rm cav}$–$L_{\rm cool}$ relation from \cite{hlavacek-larrondo_x-ray_2015} (see Figure \ref{fig:pbondi_v_pcav}).

While the correlation between cavity power and cooling luminosity is well established, the exact mechanism by which energy from AGN feedback heats the ICM remains uncertain. In some systems, expanding cavities are surrounded by shock-heated shells of gas, implying that supersonic inflation can drive shocks into the surrounding medium and locally deposit heat \citep{fabian_deep_2003}. In Section \ref{subsec:spectral_maps}, we identified a hot arc of gas along the rim bordering Cavities 1 and 2, which may represent a weak shock from supersonic bubble expansion. However, alternative interpretations—such as projection effects caused by the displacement of cooler gas—are equally plausible, and even if the arc is shock-heated, its limited spatial extent and modest Mach number suggest it contributes little to the overall energy budget required to balance cooling.

To assess the broader impact of AGN feedback, we turn to the large-scale surface brightness edges identified in Section \ref{subsec:cocoon}. Using the same sectors employed earlier to estimate the ages of the shocks in Section \ref{subsubsec:cav_heat}, we estimate the total energy carried by all three shocks to be $\sim 4 \times 10^{60}$ erg. Of this, roughly 3\% is stored in the internal energy of the cavities themselves (see Table \ref{tab:xray_cavity_measurements}). 

The corresponding shock powers, derived from the same age estimates, are $5 - 9 \times 10^{44}$ erg $s^{-1}$, roughly matching the cooling luminosity at each shock’s radius. Although the total power from shocks reaches $\sim 2 \times 10^{45}$ erg s$^{-1}$, this alone is not sufficient to fully inhibit cooling over long timescales. For example, at 50 kpc, where the radiative losses are $\sim 2 \times 10^{44}$ erg s$^{-1}$ and the local cooling time is $\sim 2$ Gyr, the total energy lost to radiation would require $\sim 10^{61}$ erg of cumulative heating. In contrast, the energy carried by the innermost shock is only $\sim 1.4 \times 10^{60}$ erg, implying that approximately 5 such shocks would be required to offset cooling at that radius over the full cooling time. Assuming a typical outburst interval of $\sim$60 Myr, as inferred from the duty cycle estimated in Section \ref{subsubsec:cav_heat}, $\sim$45 shocks could occur within 2 Gyr—well above the $\sim$5 needed in principle. However, this scenario assumes that each shock efficiently deposits its energy as heat into the ICM.

Weak shocks typically deposit only a small fraction of their energy, e.g. 5\% in NGC 5813 \citep{randall_shocks_2011}, as heat within the cooling radius. Within the innermost edge, we detect the most significant pressure and entropy increase within the $260-320^\circ$ sector. Following \citet{randall_shocks_2011}, we estimate the energy injected by the shock that directly increased the entropy of the ICM using $\Delta Q \simeq T \Delta S$. For weak shocks, the entropy increase $\Delta S$ corresponds to a fractional heat input relative to the gas thermal energy, given by $\frac{\Delta Q}{E} \sim \Delta \ln (P / \rho^\gamma)$, where $E$ is the thermal energy and $\gamma$ is the adiabatic index.

For the $\sim 50$ kpc shock, we estimate the pressure jump across each shock using the Rankine-Hugoniot relation:
\begin{equation}
\frac{P_2}{P_1} = 1 + \frac{2 \gamma}{\gamma + 1} (\mathcal{M}^2 - 1),
\end{equation}
where $\gamma = 5/3$. Using the values in Table \ref{tab:sb_edges} for the $260-320^\circ$ sector, we find that shock heating replaces $< 0.1\%$ of the local thermal energy of the gas at the shock front. Therefore, while $\sim45$ outbursts comparable in strength to the innermost shock would in principle be sufficient, the cumulative energy released over the cooling timescale still falls short of the $\sim10^{61}$~erg required for thermal balance by roughly a factor of three. As such, the shocks would need to deposit about three times as much mechanical energy as heat to fully offset radiative losses, suggesting that shock heating alone is unlikely to maintain thermal equilibrium across the cluster. This result is consistent with the broader argument that while bubble enthalpy, weak shocks, and turbulent dissipation contribute to heating the ICM, they do not completely suppress cooling, as evidenced by the persistence of multiphase gas cospatial with the coldest X-ray bright gas.

\subsection{Fueling the SMBH}
\label{subsec:agn_accretion}

The coexistence of hot, warm, and cold gas phases in the core of A2597 naturally raises the question: which phase primarily fuels the SMBH? Given the sustained AGN activity, a major and recurring fuel supply is necessary to maintain its power output. At the scale of the Bondi radius, where the SMBH’s gravitational influence dominates over thermal gas motions, fueling mechanisms are generally classified into two categories: hot fueling and cold fueling. 

Hot fueling typically refers to Bondi accretion \citep{bondi_spherically_1952}, where $\sim 1$ keV gas is directly accreted onto the SMBH. To estimate an upper limit on the Bondi accretion rate within 760 pc, the native \chandra\ resolution, we use Equation 6 from \cite{rafferty_feedback-regulated_2006}, which relates the accretion rate to the electron density $n_e$, gas temperature $T$, and black hole mass $M_{\rm BH}$ as: 
\begin{equation}
\dot{M}_{B} =  0.012 \Big(\frac{T}{\rm keV}\Big)^{-3/2} \Big(\frac{n_e}{\rm cm^{-3}}\Big)  \Big(\frac{M_{\rm BH}}{10^9 M_\odot}\Big)^2 \rm M_\odot yr^{-1}.
 \label{eqn:mass_r}
\end{equation}
To update the black hole mass estimate from \citetalias{tremblay_residual_2012}, we use the spatially resolved MUSE data presented in \cite{tremblay_galaxy-scale_2018} to adopt a stellar velocity dispersion of $\sim 290$ \kms\ within the nucleus. Using the $M_{\rm BH} - \sigma$ relation from \cite{mcconnell_two_2011} for elliptical galaxies, we calculate $M_{\rm BH} = 1.3 \times 10^9$ \msun, giving a Bondi radius of $\sim 22$ pc. 

\begin{figure}
\hspace{-2mm}
\includegraphics[width=\linewidth]{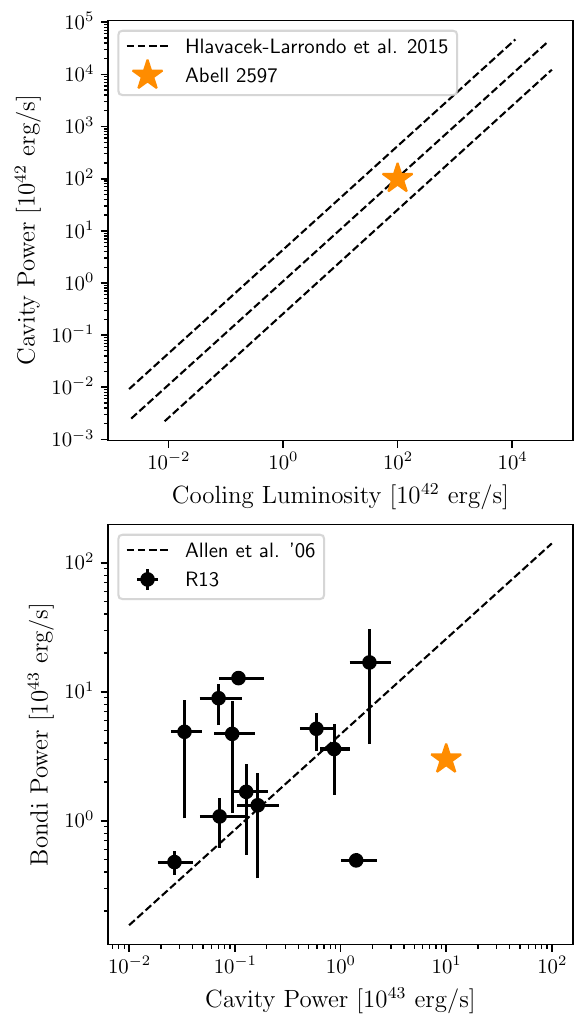}
\vspace{-2mm}
\caption{\textit{Top:} Cavity power vs. cooling luminosity for Abell 2597 (orange star), plotted alongside the best-fit relation from \citet{hlavacek-larrondo_x-ray_2015} (dashed lines). Abell 2597 follows the typical trend where AGN cavity power scales with the radiative cooling luminosity of the ICM. \textit{Bottom:} Cavity power vs. Bondi power, adapted from \citet{russell_massive_2013} and \citet{allen_relation_2006}. The Bondi power for Abell 2597 is significantly lower than the observed cavity power, suggesting Bondi accretion is not the dominant fueling mechanism.}
\label{fig:pbondi_v_pcav}
\end{figure}

Since \chandra\ cannot resolve the Bondi radius $r_B$ in A2597, we extrapolate $n_e$ and $T$ down to $r_B$ using a beta model for the density profile \citep{russell_massive_2013} and assume the temperature at the Bondi radius is approximately one-fourth of the central temperature from Table \ref{tab:chandra_profiles}. Using $n_e = 0.11\ \rm cm^{-3}$, we estimate the Bondi accretion rate at 760 pc to be $\dot{M}_{B, 760 \rm pc} \sim 5.3 \times 10^{-3}$ \msun\ yr$^{-1}$. The Eddington accretion rate, calculated using Equation 5 in \cite{rafferty_feedback-regulated_2006}  and $\epsilon = 0.1$,  results in an accretion rate of $\sim 30$ \msun yr$^{-1}$. Both the Bondi and cavity-power-derived accretion rates are sub-Eddington, consistent with \citetalias{tremblay_residual_2012}.

To explore the implications for AGN fueling, we calculate the Bondi power as  $P_{\rm Bondi} = \eta \dot{M}_{\rm Bondi} c^2$, where $\eta = 0.1$ represents the radiative efficiency. Using our estimate for the Bondi accretion rate, this yields $P_{\rm Bondi} \sim 3 \times 10^{43}$ erg s$^{-1}$.  The lower panel of Figure \ref{fig:pbondi_v_pcav} compares the Bondi power to the cavity power for A2597, along with data from \citet{russell_massive_2013}. In A2597, the Bondi power falls an order of magnitude below the estimated cavity power, suggesting that hot mode accretion alone may not provide sufficient fuel to power the observed AGN activity. This result is consistent with the findings for M84, where \chandra\ resolves the Bondi radius, yet the Bondi power is two orders of magnitude below the jet mechanical power \citep{russell_inside_2015, bambic_agn_2023}. In both clusters, assuming a lower mechanical efficiency of  $\eta \sim 10^{-2} - 10^{-3}$, as suggested by simulations \citep[e.g.,][]{sadowski_kinetic_2017}, would further reduce the Bondi power by a factor of 10-100, making it an even less viable fueling mechanism for the AGN.

Early studies of nearby elliptical galaxies \citep[e.g.,][]{allen_relation_2006}, reported a correlation between Bondi accretion power and AGN jet power, inferred from the enthalpy of jet-blown cavities. However, this correlation is subject to significant uncertainties, including projection effects and assumptions about subsonic cavity inflation. A subsequent analysis by \citet{russell_massive_2013} using a larger sample found a weaker correlation, further emphasizing that Bondi accretion alone cannot fully explain the observed AGN power.  In A2597, as in other systems, hot-mode fueling appears insufficient to sustain the AGN, indicating that additional fueling—likely from cold gas—is required.

In the cold fueling channel, thermally unstable gas cools and condenses out of the hot ICM, forming dense clouds and filaments. In Chaotic Cold Accretion (CCA) theory \citep{gaspari_chaotic_2013,gaspari_shaken_2018}, thermal instability is indicated by the condensation ratio $C \equiv t_{\rm cool}/t_{\rm eddy}$, where $t_{\rm eddy}$ is the eddy turnover time and given by \begin{equation}
t_{\rm eddy} = 2 \pi \frac{r^{2/3} L^{1/3}}{\sigma_{v, L}},
\end{equation} where $L$ is the turbulence injection scale and $\sigma_{v, L}$ is the gas velocity dispersion. 

\begin{figure*}
\includegraphics[width=\linewidth]{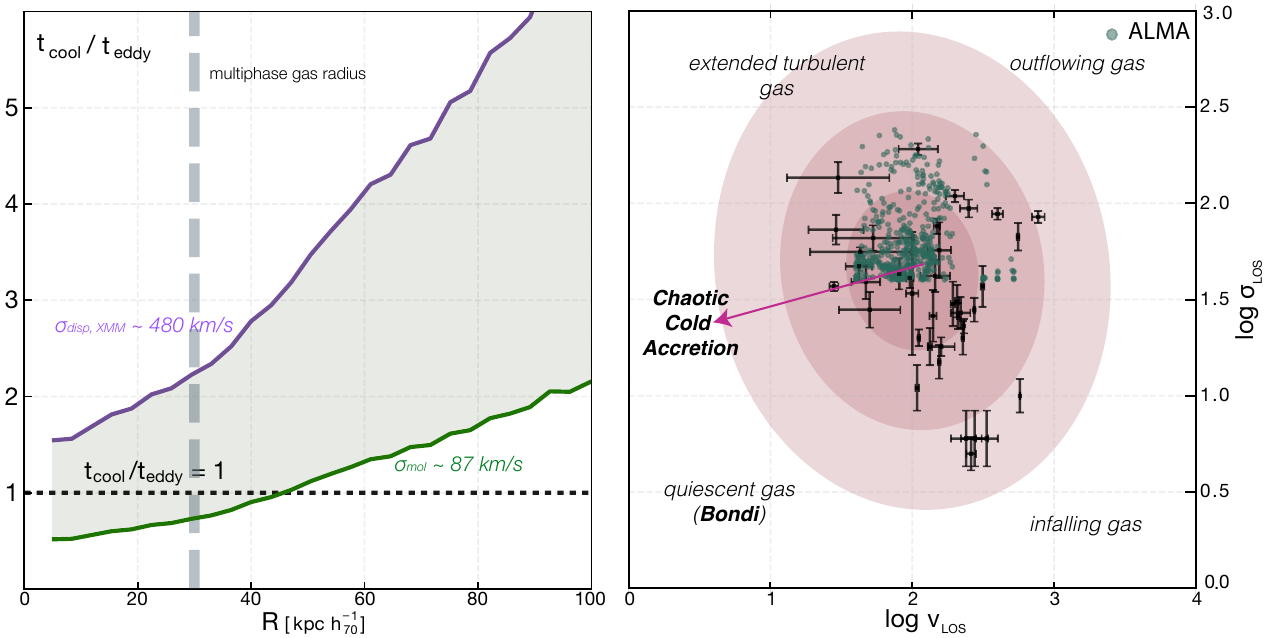}
\caption{\textit{Left:} Condensation-ratio ($C \equiv t_{\text{cool}}/t_{\text{eddy}}$) in A2597, with colors indicating different velocity dispersions used to estimate $t_{\rm eddy}$. The vertical dashed line marks the central 30 kpc, where cooler multiphase gas exists, while the horizontal dotted line indicates $t_{\text{cool}}/t_{\text{eddy}} = 1$, below which turbulent thermal instabilities are expected.   \textit{Right:} k-plot showing the kinematics of cold gas traced by archival ALMA CO(2-1) data \citep{tremblay_galaxy-scale_2018}, with velocity dispersion on the y-axis and line-of-sight velocity on the x-axis. Cold-phase points (green) represent individual spatially resolved measurements and align with the $1$–$3\sigma$ contours from pencil-beam (aperture $r \leq 25$ pc) CCA simulations \citep{gaspari_shaken_2018}, which predict $1\sigma$ asymmetric ranges of $\bar{v}_{\mathrm{los}} = 100^{+200}_{-70}$\kms for the mean line-of-sight (LOS) velocity and $\sigma_{v,\mathrm{los}} = 40^{+70}_{-30}$\kms for the line broadening. Black points show median kinematics for gas in entire massive galaxies rather than individual spaxels, including A2597 nuclear absorption features (bottom-right points). A2597’s cold-phase concentration in the central raining zone favors cold fueling (CCA) over hot fueling (Bondi accretion).}
\label{fig:kplot}
\end{figure*}

Since turbulence in the ICM is difficult to measure directly, we estimate the ICM's velocity dispersion using two independent tracers. As a lower limit, we adopt the velocity dispersion of the CO filament, $\sigma_{\rm mol, 1D}$, which has a median line-of-sight value of $\sim 50$ \kms \citep{tremblay_galaxy-scale_2018}. Correcting for three-dimensional motion, we obtain $\sigma_{mol} \approx 87$ km s$^{-1}$. For an upper limit on the ICM velocity dispersion, we take the $\sim 480 \pm 120$ \kms\ estimate from Table 4 of \citet{sanders_velocity_2013}, which derived the velocity width of the cool X-ray emitting gas from \textit{XMM-Newton} RGS spectral broadening and \chandra\ surface brightness fluctuations. Following previous studies \citep[e.g.,][]{gaspari_shaken_2018,juranova_cooling_2019,olivares_gas_2022,temi_probing_2022}, we estimate the injection length scale $L$ using the extent of the filamentary gas. Based on the \halpha\ contours in Figure \ref{fig:timescales}, we adopt $L \sim 30$ kpc, which is nicely consistent with the $\sim 22$ kpc average bubble diameter (see Table \ref{tab:xray_cavity_measurements}).

The left panel of Figure \ref{fig:kplot} shows the condensation ratio falling below unity within the central 30 kpc when using $\sigma_{\rm mol}$ and ranging between $C \sim  1-2$ with $\sigma_{\rm XMM}$. These results suggest that turbulence-driven thermal instability allows the ICM to cool rapidly over tens of kpc. If the resulting cold structures lose angular momentum through inelastic collisions, they can accrete onto the SMBH at rates exceeding those of Bondi accretion alone \citep{pizzolato_nature_2005, gaspari_chaotic_2015}. 

\begin{figure*}
\includegraphics[width=\linewidth]{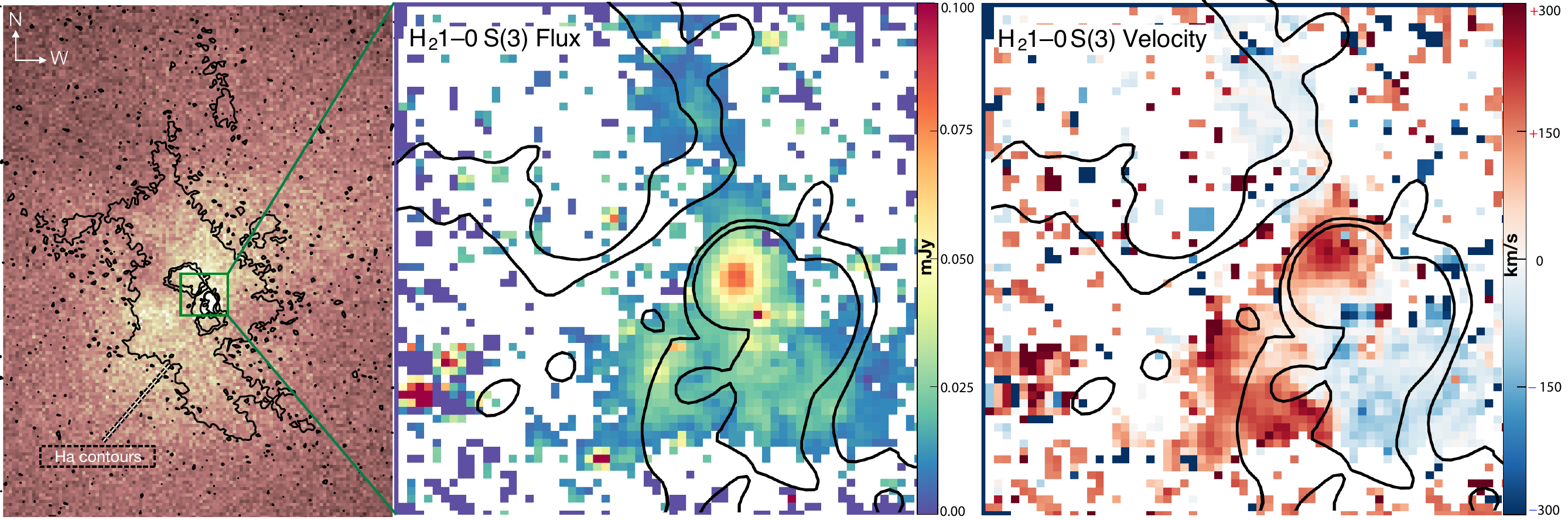}
\caption{Distribution and kinematics of the multiphase gas in the core of A2597. \textit{Left two panels:} \chandra\ X-ray map with black \halpha contours, showing warm gas draped around the rims of the southwestern cavity and northern ghost cavities. The green square marks the region shown in the new SINFONI AO maps.  \textit{Right two panels:} The SINFONI H$_2$ (1-0) S(3) flux map (middle right) shows molecular gas aligned with the edges of the northern and southern radio lobes of the 8.4 GHz radio source (black contours). The velocity map (right) shows molecular gas kinematics influenced by interactions with the radio lobes.}
\label{fig:multiphase_nuclear}
\end{figure*}

CCA predicts that turbulence enhances cold gas accretion, leading to intermittent bursts in SMBH fueling up to two orders of magnitude above the Bondi rate \citep{gaspari_raining_2017}. To better assess the kinematics of the cold gas in A2597, the right panel of Figure \ref{fig:kplot} presents the pencil-beam kinematical plot (k-plot) diagnostic from \citet{gaspari_shaken_2018}, where “pencil-beam” refers to small-aperture (less than a few arcseconds) observations that resolve localized gas kinematics. The plot shows the relationship between velocity dispersion $\sigma_{\rm los}$ (tracing turbulence) and line-of-sight velocity shifts $v_{\rm LOS}$ (tracing bulk motions) in the multiphase gas, with previously published values from massive galaxies (black points) and ALMA-detected clouds in A2597 (green circles). The cold clouds in A2597 closely follow CCA predictions, with most falling within the 1$\sigma$ contours of high-resolution CCA simulations.

While the center of the k-plot defines the strong ``raining" region, the four outer quadrants trace other potential regimes at play (see labels in Figure \ref{fig:kplot}). We observe several clouds drifting in the macro-scale weather (high turbulence, but low shifts) and a few in the opposite regime with high bulk velocities and very narrow lines, which are often tied to the infalling clouds toward the nuclear region.
The black points in the bottom-right quadrant are ALMA absorption features (against the radio AGN continuum) detected in A2597 \citep{tremblay_cold_2016}, interpreted as ``shadows'' cast by the infalling clouds. These massive ($\sim 10^5-10^6$ \msun) clouds could supply an accretion rate of $\sim 0.1$ to a few \msun yr$^{-1}$ if they are indeed falling directly towards the SMBH; the ultimate feeding rate would depend on the efficiency of angular momentum loss mechanisms (e.g., inelastic collisions and effective viscosity; \citealt{gaspari_raining_2017}). 

Conversely, the strong outflowing (top-right) and quiescent accretion quadrants (bottom-left) are not significantly filled in A2597, again consistent with the above findings that hot-mode feeding is secondary, and that A2597 does not drive ultrafast outflows \citep[e.g.,][]{laha_ionized_2021}. Observations of A2597 (and several other systems; e.g., \citealt{olivares_gas_2022,olivares_h-x-ray_2025,temi_probing_2022}), including CO absorption lines tracing molecular gas inflow \citep{rose_does_2023}, further support this `raining' picture. 

In future work, we will present the deep \textit{Chandra} data alongside new JWST MIRI and NIRSPEC observations of the nucleus of A2597. This will allow us to better constrain the gas mass budget and kinematics within the Bondi/BH influence radius, and better determine the relative contributions of cold and hot gas to the feeding process within meso-micro scales.

\subsection{The Multiphase Fountain}

Although the residual cooling rate is sufficient to supply the entire cold gas mass reservoir, it is clear that multiple rounds of AGN feedback have significantly shaped the distribution of this gas, as noted in Section 4.2 of \citealt{tremblay_galaxy-scale_2018}. The left panel of Figure \ref{fig:multiphase_nuclear} shows that the warm gas traced by \halpha is cospatial with the outer X-ray cavities, draping around the rims of the southwestern teardrop-shaped cavity, as well as the northern ghost cavities. In the nuclear region, the right panel of Figure \ref{fig:multiphase_nuclear} shows the warm molecular gas probed by the vibrational $H_2$ (1-0) transition is similarly draped around the edges of both the northern and southern lobes of the 8.4 GHz radio source, suggesting that the radio lobes are dynamically shaping the inner gas reservoir. The spatial correlation between the multiphase gas and the cavities implies two possibilities: (1) the gas has been entrained and uplifted by the jetted outflows, and/or (2) it has formed \textit{in situ} from cooling warm outflows.

\cite{tremblay_galaxy-scale_2018} primarily explored the first scenario, where pre-existing cold gas is uplifted by buoyant X-ray cavities. They demonstrated that the mechanical energy of the cavity network is sufficient to lift the entire cold molecular nebula, as the displaced hot gas mass exceeds the cold gas mass, making uplift energetically feasible based on Archimedes’ principle. Further supporting this scenario is the apparent alignment of the cold gas with the X-ray cavities in projection and the bulk velocities of the H$\alpha$ filaments ($\sim 375$ \kms), which are consistent with the terminal velocity of the rising bubbles (see Table \ref{tab:xray_cavity_measurements}). However, this alignment is only observed in projection, and without velocity measurements of the hot gas, it remains uncertain whether the cold gas is physically co-moving with the bubbles or simply coincident along our viewing angle. Future XRISM observations will be crucial in resolving this question, particularly since the kinematics of the extended multiphase nebula in Abell 2597 are mapped in far greater detail than in other cool-core clusters like Perseus and M87.

Alternatively, some hydrodynamical simulations suggest that AGN outflows can trigger in-situ cooling, where filament velocities are primarily governed by bulk bubble motions rather than turbulence \citep[e.g.,][]{qiu_formation_2020, zhang_bubble-driven_2022}. Since the buoyant rise time of the bubbles ($\sim 10^7$ yr) is comparable to the cooling time within the region where cold filaments are observed, some fraction of the cold gas may have condensed from the warm phase while being uplifted. However, the k-plot in Figure \ref{fig:kplot} shows no strong bulk motions in the cold gas, suggesting that compressional condensation through outflows plays a secondary role compared to global condensation driven by the ICM weather. 

Regardless of whether the cold gas is primarily uplifted or formed in situ, its observed distribution remains closely tied to the structure of the hot ICM (see left panel of Figure \ref{fig:multiphase_nuclear}). A recent study by \cite{olivares_h-x-ray_2025} found a strong correlation between the X-ray and H$\alpha$ surface brightness of filaments across multiple clusters, suggesting a direct connection between the hot and warm gas phases. In Abell 2597, the X-ray to H$\alpha$ surface brightness ratio remains nearly constant at around 3, slightly below the sample average of 4, regardless of the distance to the AGN. This consistency indicates that local processes—such as turbulent mixing layers, shocks, or reprocessing of extreme ultraviolet and X-ray radiation from the cooling ICM—play a dominant role in regulating the excitation of the filaments rather than AGN proximity alone. These results are also consistent with predictions from CCA models, which our analysis of cold filaments in Section \ref{subsec:agn_accretion} independently supports.

\section{The Origin of the X-ray Channel}
\label{subsec:origin_channel}

With a projected width of $\sim$8 kpc and an estimated line-of-sight depth of $\sim$9 kpc, the elongated X-ray surface brightness “channel” identified in Figure \ref{fig:xray_features} is likely a sheet of lower-density gas. Similar channels have also been observed in other systems, such as the merging cluster Abell 520 and the sloshing cool core of Abell 2142 \citep{wang_merging_2016, wang_deep_2018}. While such structures might theoretically result from two opposing cold fronts, this scenario is unlikely for A2597, as it was for A520 and A2142. Instead, these features are more plausibly interpreted as plasma depletion layers (PDLs), a phenomenon observed near planets \citep[e.g.,][]{oieroset_magnetic_2004} and reproduced in magnetohydrodynamic simulations of galaxy clusters \citep{zuhone_sloshing_2011}. In the latter simulations, PDLs form when sloshing motions amplify magnetic fields through shear flows \citep{keshet_dynamics_2010}, reducing thermal pressure and creating low-density channels (see Figure 23 of \cite{zuhone_sloshing_2011}). 

In Abell 520, the observed PDL aligned with the direction of a secondary subcluster merger \citep{wang_merging_2016}. Although Abell 2597 has historically been classified as a relaxed cluster, it is possible that the surface brightness edges identified are cold fronts, as discussed in Section \ref{subsec:agn_history}. This would imply residual sloshing triggered by a recent minor merger, and provide a PDL formation scenario consistent with that of Abell 520. If instead the edges are shock fronts—as remains our favored interpretation—simulations by \cite{zuhone_sloshing_2011} show that residual sloshing motions from past major mergers can still generate PDLs. Since galaxy clusters are expected to form a dominant halo by $z\sim 2$ ($t_H \sim 3.4$ billion years) \citep{boylan-kolchin_resolving_2009} and subsequently grow to become the nearby clusters we see today through mergers, lingering sloshing motions in A2597 may have played a role in forming the channel.

PDLs often form near cold fronts, given that cold fronts mark regions with particularly strong bulk motions. In Abell 520, the PDL is observed adjacent to a cold front. If we interpret the three detected surface brightness edges in A2597 as cold fronts, the channel’s proximity to the inner edge would resemble the configuration in Abell 520. However, PDLs do not necessarily require an associated cold front to form. Preliminary results from Bellomi et al. (in prep.) show that PDLs form in seemingly random locations within the TNG Cluster simulations \citep{nelson_introducing_2024, truong_x-ray-inferred_2024}. 

Although we favor interpreting the channel as a PDL, its presence does not necessarily imply strong magnetic fields. Simulations by \citet{chadayammuri_turbulent_2022} found that the ``channel” feature observed in Abell 520 appeared both in regions with strong magnetic fields, and in purely hydrodynamic simulations without magnetic fields (see their Figures 7 and 8). Simulations by \citet{gaspari_relation_2014} (Figure 4) also show that comparable low-density structures can emerge purely due to hydrodynamics, with turbulent eddies naturally creating alternating regions of high and low density. Therefore, without resolved measurements of the magnetic field strength in A2597, the exact origin of the channel remains uncertain.

\section{Summary}

In this paper, we presented deep ($\sim 600$ks) \chandra\ X-ray observations of the cool core galaxy cluster Abell 2597, complemented by archival GMRT radio data and SINFONI near-infrared observations, to study the interplay between AGN feedback, radiative cooling, and black hole accretion. The new X-ray data provide the most detailed view to date of the cluster's ICM, with our results summarized as follows:

\begin{enumerate}
    \item \textit{Identified structures:} We identified three new X-ray cavities, increasing the total number in the cluster core to seven. The cavities are roughly aligned along the same southwest-northeast direction as the extended radio emission. Metal-enriched gas is also preferentially distributed along this axis, supporting a scenario where AGN outflows redistribute metals into the surrounding ICM. Beyond the central 50 kpc, we detect three subtle surface brightness edges with corresponding jumps in temperature, pressure, and entropy, which we interpret as potential shock fronts. Within the innermost edge, we also identify an elongated X-ray surface brightness deficit, or ``channel,” which we interpret as a plasma depletion layer.

    \item \textit{Inefficient Energy Injection from AGN Feedback:} The cavities and potential shock fronts inject energy into the ICM at an estimated rate of $\sim 10^{44}$ erg s$^{-1}$ - $\sim 10^{45}$ erg s$^{-1}$, comparable to / exceeding the cluster’s cooling luminosity ($L_{\rm cool} \sim 10^{44}$ erg s$^{-1}$). However, this energy appears insufficient to fully counteract radiative cooling within $r_{\rm cool} \sim 60$ kpc, with $< 1\%$ of the local thermal energy replaced by shock heating. The residual X-ray and UV cooling rate is just enough to fill the billion solar mass cold molecular gas reservoir near the core of the cluster, providing further evidence that AGN feedback alone does not entirely suppress cooling flows, as expected in a healthy self-regulating feedback loop.

    \item \textit{Duty Cycle of the AGN:} The presence of multiple cavities and subtle surface brightness edges (which we interpret as weak shocks) at various distances from the AGN suggests that they may have formed over several discrete AGN outbursts.  If so, the inferred interval between events is $\sim 10^7$ years—a timescale remarkably consistent across cool-core systems, from nearby groups to distant clusters. However, it remains possible that all the observed cavities were produced during the most recent AGN outburst, rather than from multiple discrete events, with the bubbles fragmenting due to Rayleigh-Taylor instabilities as they rose. Future high-resolution radio observations and spectral aging measurements will be necessary to distinguish between these scenarios.

    \item \textit{Feeding the AGN:} With an estimated Bondi power of $P_{\rm Bondi} \sim 2 \times 10^{43}$ erg s$^{-1}$, an order of magnitude lower than the cavity power, hot mode accretion alone likely cannot sustain the AGN. A2597 ALMA observations of cold molecular clouds show kinematics consistent with turbulent condensation, supporting the chaotic cold accretion scenario as the dominant feeding mechanism.  

    \item \textit{The X-ray Channel:} The ``channel" spans 57 kpc in length and 8 kpc in width, located near the inner surface brightness edge. It may be a plasma depletion layer, a low-density gas structure formed by sloshing-induced magnetic field amplification from a past  merger. If the surface brightness edges are interpreted as cold fronts rather than shocks, this would support a sloshing origin, consistent with scenarios proposed for similar channels in Abell 520 and Abell 2142. However, PDL formation does not always require strong magnetic fields. Without resolved magnetic field measurements, the channel’s origin remains uncertain.
    
\end{enumerate}

While our observations provide a detailed view of the interplay between AGN feedback and cooling in Abell 2597, several open questions concerning the precise mechanisms driving gas and metal uplift, the contributions of hot-phase versus cold-phase SMBH fueling, and the role of magnetic fields in shaping ICM structures require further exploration. Pursuing deeper, higher-resolution radio observations with the uGMRT and X-ray spectroscopy XRISM will be crucial for addressing these uncertainties and advancing our understanding of feedback and accretion in cool-core clusters.

\vspace{5mm}
\hspace{15mm} \noindent \textsc{ACKNOWLEDGMENTS}

We thank the referee for a careful reading of our manuscript and for providing feedback that improved this paper significantly.

The scientific results reported in this paper are based on observations from multiple ground- and space-based observatories, including the \textit{Chandra} X-ray Observatory, the Giant Metrewave Radio Telescope, the upgraded Giant Metrewave Radio Telescope, and the SINFONI near-infrared spectrograph on the Very Large Telescope. PK acknowledges the support of the Department of Atomic Energy, Government of India, under the project 12-R\&D-TFR-5.02-0700.

\textit{Chandra} data were obtained through programs supported by the Chandra X-ray Center, which is operated by the Smithsonian Astrophysical Observatory for NASA under contract NAS8-03060. The GMRT and uGMRT data were accessed through the National Centre for Radio Astrophysics (NCRA), Tata Institute of Fundamental Research (TIFR), located in Pune, India. GMRT is operated by NCRA-TIFR, and we acknowledge the significant contributions of its engineering and technical teams in making the observations possible. We also use SINFONI observations obtained at the Very Large Telescope of the European Southern Observatory, Paranal, Chile, under program ID 0102.A-0463(A). These data were critical for mapping the distribution and kinematics of warm molecular gas in the cluster core. We acknowledge the efforts of the teams behind these facilities for enabling the collection of the high-quality data used in this work. 

\textit{Facilities: ALMA, CXO, VLT, HST, GMRT, VLA}

\textit{Software:} \textsc{Astropy} \citet{robitaille_astropy_2013, collaboration_astropy_2018, collaboration_astropy_2022}, \textsc{CASA} \citep{mcmullin_casa_2007}, \textsc{CIAO} \citep{fruscione_ciao_2006}, \textsc{IPython} \citep{perez_ipython_2007}, \textsc{Matplotlib} \citep{caswell_matplotlibmatplotlib_2022}, \textsc{Numpy} \citep{van_der_walt_numpy_2011}, \textsc{pyproffit} \citep{eckert_low-scatter_2020}, \textsc{PySpecKit} \citep{ginsburg_pyspeckit_2011}, \textsc{scipy} \citep{virtanen_scipy_2020}

\bibliography{a2597_bib_filtered}
\bibliographystyle{aasjournal}

\appendix

\section{Single Power-Law Fits to Candidate Edges}
\label{app:spw_edges}

The surface brightness profiles for each candidate edge were also fit with single power-law models for comparison. The results are reported in Table \ref{tab:app_sp_edges}.

\begin{table*}[!h]  
\hspace*{-3cm}
\centering  
\begin{tabular}{lcccccc}  
\hline  
\hline  
Edge & Sector & $r_J$ & $\alpha$ & $\chi^2$ & $\mathrm{ d.o.f}$ & $\bar\chi_\nu^2$ \\
 & $(^\circ)- (^\circ)$ & (kpc)  &  &  & \\
\hline
1 & 20-80 & 49 & 1.99 $\pm$ 0.02  & 118.25 & 19 & 6.22 \\
 & 80-140 & 47 & 2.28 $\pm$ 0.02  & 118.22 & 19 & 6.22 \\
 & 140-200 & 45 & 2.09 $\pm$ 0.02  & 178.72 & 19 & 9.41 \\
 & 200-260 & 39 & 1.87 $\pm$ 0.02  & 133.80 & 17 & 7.87 \\
 & 260-320 & 49 & 2.06 $\pm$ 0.02  & 95.82 & 22 & 4.36 \\
 & 320-380 & 43 & 2.00 $\pm$ 0.02  & 219.53 & 20 & 10.98 \\
 \hline
 2  & 20-80 & 96 & 2.07 $\pm$ 0.04  & 20.62 & 17 & 1.21 \\
 & 80-200 & 90 & 2.23 $\pm$ 0.07  & 38.01 & 32 & 1.19 \\
 & 200-260 & 96 & 2.27 $\pm$ 0.02  & 48.63 & 30 & 1.62 \\
 & 260-380 & 88 & 2.35 $\pm$ 0.02  & 48.32 & 31 & 1.56 \\
 \hline
 3 & 20-80 & 152 & 2.55 $\pm$ 0.12  & 25.09 & 28 & 0.90 \\
 & 80-200 & 144 & 2.35 $\pm$ 0.04  & 30.36 & 19 & 1.60 \\
 & 260-380 & 130 & 2.51 $\pm$ 0.03  & 33.50 & 24 & 1.40 \\
\hline  
\end{tabular}  
\caption{Best-fitting parameters of the single power-law models for the candidate surface brightness edges in Abell 2597, corresponding to the sectors shown in Figure~\ref{fig:cs_profiles}. Columns list the azimuthal range of each sector, projected radius from the center of the profile, $\chi^2$, degrees of freedom (d.o.f.), and reduced $\chi^2$ ($\bar \chi^2_\nu$). The model provides a poor fit for Edge 1 ($\bar\chi^2_\nu >> 1$), supporting its classification as a genuine edge. Conversely, the good fits for Edges 2 and 3 ($\bar\chi^2_\nu \sim 1$) suggest they are more subtle surface brightness breaks, making their interpretation as true edges less certain.}
\label{tab:app_sp_edges}  
\vspace{-2mm}
\end{table*}  

\end{document}